\documentclass[twocolumn,pra,aps,superscriptaddress]{revtex4}
\usepackage{subfigure}
\usepackage{psfrag,graphicx}
\usepackage{pict2e}
\usepackage{dcolumn}
\usepackage{amsmath,amssymb}
\usepackage{bm}
\usepackage{color}
\usepackage{latexsym}
\usepackage{epstopdf}
\usepackage{color}
\usepackage[english]{babel}
\usepackage{amsfonts}
\usepackage{bm}
\usepackage{natbib}
\usepackage{cancel}
\usepackage{appendix}
\usepackage{bbold}
\usepackage{placeins}
\usepackage{appendix}
\usepackage{enumerate}
\definecolor{mygrey}{gray}{0.35}
\definecolor{myblue}{rgb}{0.2,0.2,0.8}
\definecolor{myzard}{cmyk}{0,0,0.05,0}
\definecolor{mywhite}{rgb}{1,1,1}
\definecolor{mywhite}{rgb}{1,1,1}
\definecolor{myred}{rgb}{1,0.,0.3}
\usepackage[colorlinks=true,citecolor=myblue,linkcolor=myred]{hyperref}
  \newcommand\underrel[2]{\mathrel{\mathop{#2}\limits_{#1}}}
\def\cc{{\rm c}}
\def\dd{{\rm d}}

\def\pp{{\rm p}}

\begin{document}
\title{Topological Amplification in Photonic Lattices}

\author{Diego Porras}
\email{d.porras@iff.csic.es}
\affiliation{Instituto de F\'{\i}sica Fundamental IFF-CSIC, Calle Serrano 113b, Madrid 28006, Spain}
\affiliation{Department of Physics and Astronomy, University of Sussex, Falmer, Brighton BN1 9QH, UK}

\author{Samuel Fern\'andez-Lorenzo}
\affiliation{Department of Physics and Astronomy, University of Sussex, Falmer, Brighton BN1 9QH, UK}

\date{\today}

\begin{abstract}
We characterize topological phases in photonic lattices
by unveiling a formal equivalence between the singular value decomposition of the non-Hermitian coupling matrix and the diagonalization of an effective Hamiltonian. 
Our theory reveals a relation between topological insulators and directional amplifiers. 
We exemplify our ideas with an array of photonic cavities 
which can be mapped into an AIII topological insulator.
We investigate stability properties and prove the existence of stable topologically non-trivial steady-state phases.
Finally, we show numerically that the topological amplification process is robust against disorder in the lattice parameters.
\end{abstract}

\maketitle
\setcounter{tocdepth}{2}
\begingroup
\hypersetup{linkcolor=black}
\endgroup

\textit{ Introduction.--} 
Topological photonics builds on ideas from electronic band theory \cite{ozawa18}, such as the classification of topological phases based on symmetries \cite{Schnyder08,ryu10,Bansil16,Chiu16}.
After Haldane and Raghu's pioneering work \cite{Haldane08}, early realizations of topological phases were implemented in photonic spin Hall systems  \cite{Hafezi11}.
In the last years topological phases have been investigated in photonic lattices by breaking time-reversal symmetry with magnetic fields  
\cite{Wang09,koch10,Anderson16,Lu2016,Owens18}
or periodic drivings \cite{fang12,Peropadre13,Rechtsman13,roushan17,sounas17,mukherjee18}. 
Analogous ideas have been explored in
optomechanical systems \cite{Hafezi12,Schmidt15,Ruesink16,Shen16,bernier18pra} or even in purely vibronic or mechanical systems \cite{Bermudez11prl,Bermudez12,Susstrunk15}, as well as in spin-cavity setups \cite{harder17prb,zhang17natcomm}.

Photonic lattices present distinctive features with respect to their electronic counterparts like dissipative decay or pumping (loss/gain), as well as coherent driving \cite{houck12natphys,Hartmann16}.
The breaking of time-reversal invariance that is a typical ingredient of topological phases leads to non-reciprocal photon transport \cite{Longhi17} and topological quantum fluctuations \cite{peano16prx}.
Loss/gain in photonic lattices leads to non-Hermitian coupling matrices in which 
the direct application of topological insulator theory is highly non-trivial.
Theoretical works have shown zero-energy edge states in non-Hermitian systems 
\cite{Esaki11,Schomerus13,Lee16,Leykam17}, and even extended topological band theory to non-Hermitian lattices \cite{Gong18,shen18}. 
Theoretical work has so far focused on topological properties of eigenvectors of non-Hermitian matrices. Recent experiments have detected photonic zero edge modes in the transmission properties of photonic lattices with a chiral symmetry originated by a bi-lattice structure \cite{Poli15,Zeuner15,Pan18,hadad18}, and shown non-reciprocal transmission induced by a synthetic magnetic field in an optomechanical system \cite{fang17}.

In this article we present a novel approach to the study of topological photonic phases that focuses on the singular value decomposition (s.v.d) of the non-Hermitian coupling matrix, $H$.
Our approach establishes a link between the existence of non-trivial topological phases and the amplification of a coherent input signal.
We show that the s.v.d of $H$ is formally equivalent to the diagonalization of an effective Hamiltonian, $\cal H$, which posses an intrinsic chiral symmetry and potentially also a set of topological protected edge states. The latter govern the response to a coherent drive and they lead to an exponential amplification effect.
Our formalism allows us to exploit the existing classification of Hermitian lattices into symmetry classes
\cite{Schnyder08,ryu10,Bansil16,Chiu16} and predict parameter regimes in which a photonic lattice acts as an amplifier.
We present an example consisting of an array of coupled photonic cavities and we find topologically non-trivial phases that are stable over a wide range or parameters. 
Numerical calculations show that our scheme is topologically protected against disorder. Finally we discuss a physical implementation of our ideas in a photonic lattice subject to periodic drivings.

\textit{Dissipative photonic lattice.--}
We consider a dissipative lattice of local photonic modes with annihilation and creation operators
$a_j$ and $a^\dagger_j$, respectively, whose density matrix operator dynamics is described by  ($\hbar = 1$)
\begin{eqnarray}
{\cal L}(\rho) &=& 
\sum_{j,l} \Gamma^{(\rm d)}_{j l} 
\left( 2 a_j \rho a^\dagger_l - a^\dagger_l a_j \rho - \rho a^\dagger_l a_j \right)
\nonumber \\
&+& 
\sum_{j,l} \Gamma^{(\rm p)}_{j l} 
\left( 2 a^\dagger_j \rho a_l - a_l a^\dagger_j \rho - \rho a_l a^\dagger_j \right)
\nonumber \\
&-&
i  [ \sum_{j,l} G_{j l} \ a^\dagger_j a_l , \rho ] 
- i  [ \sum_j \left( \epsilon^*_j a_j + \epsilon_j a^\dagger_j  \right), \rho ] .
\label{photonic.lattice}
\end{eqnarray}
$\Gamma^{\rm (d)}_{j l}$ and $\Gamma^{\rm (p)}_{j l}$ are Hermitian matrices that describe collective photon losses and incoherent pumping, respectively. Diagonal terms, $\Gamma^{\rm (d)}_{j j}$ and $\Gamma^{\rm (p)}_{j j}$, appear naturally because of local photon losses or gain, being the latter induced by an active medium or illumination with an incoherent source. Non-diagonal dissipative terms ($j\neq l$) can be controlled by using additional degrees of freedom (for example auxiliary modes \cite{Hafezi12,Ruesink16,Shen16,bernier18pra}). 
$G_{j l}$ are coherent couplings describing the tunneling of photons between cavities.
  Finally, the last term describes a resonant coherent drive with a site-dependent complex amplitude, $\epsilon_j$. The latter can be induced by direct illumination of the photonic lattice with a coherent field. 

We define the non-Hermitian matrix
$H = \Gamma - i G$, where 
$\Gamma_{jl} = \Gamma^{\rm (p)}_{jl} - \Gamma^{\rm (d)}_{lj}$, which allows us to express the evolution of the coherences, $\alpha_j = \langle a_j \rangle$,
\begin{equation}
\dot{\alpha}_j = \sum_{l=1}^N H_{j l}  \alpha_l  + \epsilon'_j ,
\label{coherences}
\end{equation}
with $\epsilon'_j = i  \epsilon_j$.
Eq. \eqref{coherences} is a closed set of linear equations that exactly describes the evolution of field coherences in the lattice-
The steady-state solution, $\alpha^{\rm ss}_j$, is found by solving 
$\dot{\alpha}_j = 0$, and it can be expressed in terms of the s.v.d., $H = U S V^\dagger$, 
where $U$ and $V$ are unitary matrices and $S$ is a diagonal matrix, 
$S_{n m} = s_n \delta_{n m}$, with $s_n \geq 0$, 
\begin{equation}
\alpha^{\rm ss}_j = - \sum_{n,l} V_{j n} s^{-1}_n U^{*}_{l n} \epsilon'_l .
\label{steady.state}
\end{equation}
Small values of $s_n$ give a larger contribution to $\alpha^{\rm ss}_j$.

\textit{Mapping to an effective Hamiltonian.--} 
Our work relies on the observation that the s.v.d of $H$ is equivalent to the diagonalization of an effective Hamiltonian, $\cal H$, 
\begin{equation}
{\cal H}  
= H \otimes \sigma^+ + H^\dagger \otimes \sigma^- ,
\label{H}
\end{equation}
where we have introduced ladder spin operators acting on an auxiliary spin-$1/2$, 
$\{ | \! \uparrow \rangle, |\!\downarrow \rangle  \}$.
We define singular vectors $u^{(n)}$, $v^{(n)}$, corresponding to the columns of $U$ and $V$, 
$u^{(n)}_{j} = U_{j n}$, $v^{(n)}_{j} = V_{j n}$,
\begin{equation}
{\cal H} 
\left( u^{(n)} \! \otimes \! | \! \uparrow \rangle 
\pm  
v^{(n)} \! \otimes \! | \! \downarrow \rangle \right) = \pm s_n 
\left( u^{(n)} \! \otimes \! | \! \uparrow \rangle 
\pm  
v^{(n)} \! \otimes \! | \! \downarrow \rangle \right).
\end{equation}
The eigenvalues of $\cal H$ come in pairs, $\pm s_n$, due to the chiral symmetry,
\begin{equation}
(\mathbb{1}\otimes \sigma_z) {\cal H} (\mathbb{1}\otimes \sigma_z) = - {\cal H},
\label{chiral.symmetry}
\end{equation}
which exists by the very definition of $\cal H$, independently of the physical symmetries of the lattice.  
The mapping $H \to {\cal  H}$ allows us to use the theoretical machinery of topological band theory (t.b.t.) \cite{Schnyder08,ryu10,Bansil16,Chiu16} and classify topological steady-states in translational invariant lattices.  We consider periodic boundary solutions and express ${\cal H}$ in a plane-wave basis,
%
\begin{equation}
{\cal H}(\vec{k}) =  \Gamma(\vec{k})  \sigma_x 
+ G(\vec{k}) \sigma_y,
\label{H.eff}
\end{equation}
where $\Gamma(\vec{k})$ and $G(\vec{k})$ are real functions due to the hermiticity of the coupling matrices.
t.b.t. relies on symmetry operators 
$T$ (time-reversal) and $C$ (charge conjugation), written like
$T = U_T K$, $C = U_C K$, where $U_T$, $U_C$ are unitary matrices and $K$ is the complex conjugation operator
($K^2 = 1$, $K i K = -i$). 
Condition $U_T U_C \propto \sigma_z$ must be fulfilled to account for the chiral symmetry expressed in Eq. \eqref{chiral.symmetry}.
Time-reversal and/or charge conjugation symmetries are fulfilled if there exist unitary matrices $U_T$, $U_C$, such that 
$T {\cal H}(\vec{k}) T^{-1} = {\cal H}(-\vec{k})$ and/or $C {\cal H}(\vec{k}) C^{-1} = -{\cal H}(-\vec{k})$, respectively. 
We find the following possible symmetry classes  \cite{ryu10}:
\begin{enumerate}[(i)]
\item 
$\Gamma(\vec{k})^2 +  G(\vec{k})^2 \neq \Gamma(-\vec{k})^2 +  G(-\vec{k})^2$  
$\to$ AIII class (no $T$, $C$ symmetry).
\item 
Vectors $(\Gamma(\vec{k}),-G(\vec{k}))$ and 
$(\Gamma(-\vec{k}),G(-\vec{k}))$ are related by a rotation with angle $\theta$ on the $x$-$y$ plane $\to$ BDI class ($T^2 = C^2 = 1$) with $U_T = \exp(i \sigma_z \theta/2)$, $U_C = \exp(i \sigma_z (\theta+ \pi)/2)$.
\item  $\Gamma(\vec{k}) = \Gamma(-\vec{k})$, $G(\vec{k}) = G(-\vec{k})$ $\to$ CI class ($T^2 = 1$, $C^2 = -1$) with $U_T = \sigma_x$, $U_C = \sigma_y$.
This is the case of real couplings matrices, $\Gamma_{jl}=\Gamma_{lj}$, $G_{jl}=G_{lj}$.
\item $\Gamma(\vec{k}) = - \Gamma(-\vec{k})$, $G(\vec{k}) = - G(-\vec{k})$ $\to$ DIII class ($T^2 = -1$, $C^2 = 1$) with $U_T = \sigma_y$, $U_C = \sigma_x$.
\end{enumerate}
This classification allows us to predict the existence or not of edge states. Whereas the chiral symmetry in Eq. \eqref{chiral.symmetry} is always fulfilled by construction, the physical symmetries of the coupling matrices $\Gamma$, $G$, determine the symmetry class above. For example, if we have real dissipative couplings $\Gamma_{jl}$, but complex photon tunneling terms, we get $G(\vec{k}) \neq G(-\vec{k})$, such that the system falls into the AIII symmetry class.

\textit{Edge singular vectors and amplification.--}
Non-trivial topological properties of $\cal H$ have dramatic consequences in the steady-state. 
In particular, t.b.t predicts the existence of zero-energy eigenstates of $\cal H$ in non-trivial topological phases (see for example Ref. \cite{ryu02}), which in turn implies the appearance of
zero-singular values, $s_{n_{\rm E}}$ that are separated by a gap from the bulk singular values, $s_{n \neq n_{\rm E}}$. 
The characterization of topological insulators in terms of symmetry classes (see \cite{Schnyder08,ryu10,Bansil16,Chiu16}) can be used to predict the existence of those edge states. 
From t.b.t. applied to $\cal H$, we also expect the emergence of right/left {\it edge} singular vectors, $u_j^{(n_{\rm E})}$/$v_j^{(n_{\rm E})}$, whose amplitude is localized at the edges of the lattice.  

We assume for simplicity that there is a single zero-singular value $n_{\rm E} = N$, corresponding to a single zero-energy state of ${\cal H}$ (as is the case in the one-dimensional model below).  
In a finite size lattice, this typically implies that $s_N \propto e^{-L/\chi}$, that is, the edge singular value decays exponentially with the length of the system, $L$, divided by a typical length, $\chi$.
The sum in Eq. (\ref{steady.state}) can thus be approximated by
$\alpha_j^{\rm ss} \approx - \sum_{l} v^{(N)}_j s^{-1}_N {u^{(N)}_l}^* \epsilon'_l$.
That expression can be simplified further in translationally invariant systems, in which the parity inversion operator, $\Pi$, fulfills that $\Pi H \Pi = H^{\rm T}$. In this case we find the condition $V = \Pi U^*$, which leads to
\begin{equation}
\alpha^{\rm ss}_j = -  \Pi {u^{(N)}_{j} }^* s^{-1}_N \sum_l {u^{(N)}_{l}}^* \epsilon'_l .
\label{amplification}
\end{equation}
We have arrived to our main result: the existence of edge states of $\cal H$ leads to the amplification of a coherent drive.
Eq. \eqref{amplification} implies that $\alpha^{\rm ss}_j$ is amplified by a factor $1/s_{N} \propto e^{L/\chi}$, and it is also proportional to the overlap between $\epsilon_j'$ and  $u_l^{(N)}$. Furthermore, field coherences $\alpha^{\rm ss}_j$ are distributed following the parity inverted singular edge-state vector, $\Pi u_j^{(N)}$.
This implies that amplification is a directional process triggered by a coherent drive in one of the system's edges and leading to large values of the field in the opposite edge.
By measuring the coherent component of the field in the photonic cavity, for example by means of homodyne detection methods, the existence of singular edge-state can be experimentally proved.

\begin{figure}[h!]
	\includegraphics[width=0.45\textwidth]{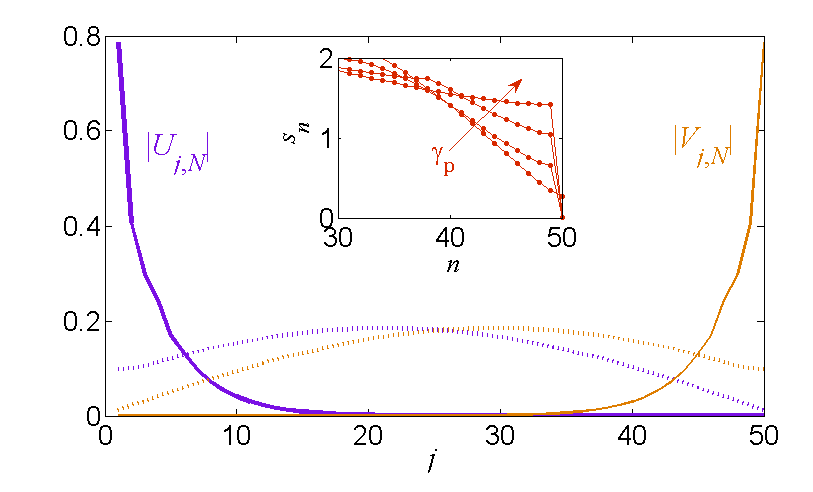}
	\caption{Purple/orange lines are the absolute value of the $n = N$ right/left singular vector of the 1D model \eqref{1D.couplings} with  $t_\cc = t_\dd = 1$, $\phi = \pi/3$, $N = 50$. Dotted lines: $\gamma_\pp = 0$ (topologically trivial case). Continuous lines: $\gamma_\pp = 1$ (topologically non-trivial case).
		Inset: Singular values for $\gamma_\pp = $ 0, 1, 1.5, 2., showing the emergence of a zero singular value state at $n_{\rm E} = N$.}
	\label{fig:fig1} 
\end{figure}

\textit{One-dimensional example: non-reciprocal photonic chain.--}
We consider an array of cavities with nearest-neighbour dissipative couplings leading to
\begin{eqnarray}
\Gamma_{j l} &=& (\gamma_{\rm p} - 2 t_\dd ) \delta_{j l} + t_{\dd} \delta_{l,j+1} + t_{\dd} \delta_{l,j-1},  \nonumber \\
G_{j l} &=&  \ t_{\cc} e^{i \phi} \delta_{l,j+1} 
+ \ t_{\cc} e^{-i \phi} \delta_{l,j-1} .
\label{1D.couplings}
\end{eqnarray}
This photonic lattice is related to the Hatano-Nelson model \cite{Hatano96,Longhi15}. Dissipative terms with rate $t_\dd$ couple nearest neightbors. The diagonal element $\Gamma = \gamma_\pp - 2 t_\dd$ is parametrized in terms of $\gamma_\pp$, which determines the net incoherent pumping of photons into the lattice. Finally, coherent couplings couple nearest-neighbor sites with complex tunneling phase, $\phi$.
The effective Hamiltonian in the plane-wave basis is
\begin{eqnarray}
{\cal H}(k) = \left(\gamma_{\rm p} \! - \! 2 t_\dd \! +  \! 2 t_\dd \cos(k) \right) \sigma_x
+ 2 t_\cc \cos(k \! + \! \phi) \sigma_y .
\label{eq:H1D}
\end{eqnarray}
The cases $\phi = 0, \pi$ belong to the CI class. 
The generic case with $\phi \neq 0$ belongs to the AIII class and we can expect non-trivial topological phases to appear \cite{Chiu16}. 
We characterize the properties of Hamiltonian \eqref{eq:H1D} by using the winding number, $\nu$, as a topological invariant \cite{asboth16}.
For this we write ${\cal H}(k) = \Gamma(k) \sigma_x + G(k) \sigma_y$, such that
$\nu = 1$ (non-trivial topological phase) if the ellipse formed by the two-dimensional vector 
$(\Gamma(k), G(k))$ encompasses the origin as $k$ varies from $0$ to $2 \pi$, and $\nu = 0$ otherwise.
Without loss of generality we can assume $0 \leq \phi < \pi$.
The condition for $\nu = 1$ reads
\begin{equation}
2 t_\dd \left( 1 - \sin \phi \right) < \gamma_\pp  < 2 t_\dd  \left( 1 + \sin \phi \right) , 
\label{eq:condition_topology}
\end{equation}
and $t_\cc \neq 0$. Since the case $\phi =0$ and/or $\gamma_\pp = 0$ do not admit non-trivial solutions, our model requires complex photon tunneling couplings together with incoherent pumping in the original photonic lattice.
A numerical calculation confirms the appearance of zero singular value modes, see Fig. \ref{fig:fig1}.

We focus now in the case $\phi = \pi/2$, $t_\dd = t_\cc$, which can be mapped into the Su-Schriefer-Heeger (SSH) model \cite{Heeger88}. Eq \eqref{eq:condition_topology} implies that non-trivial topological phases exist if $0 < \gamma_\pp < 4 t_\dd$.
Here we can analytically find expressions for the edge states wave-functions which can be written in terms of the localization length \cite{Venuti07}, 
$\xi^{-1} = - \ln| 1 - \frac{\gamma_\pp}{2 t_\dd} | $.
In the following we assume the limit $N \gg \xi \gg 1$, to simplify the discussion, such that
the edge singular vector is 
$u_j^{(N)} = \sqrt{2/\xi} e^{-j/\xi}$, with singular value
$s_N = 2 \gamma_\pp e^{-N/\xi}$, leading to
\begin{equation}
\alpha^{\rm ss}_j = - \frac{1}{\gamma_\pp \xi} e^{N/\xi}   e^{-(N+1-j)/\xi} \sum_l e^{-l/\xi} \epsilon'_l .
\label{amplify}
\end{equation}
The signal is thus directionally amplified by coherently driving the left end of the chain such that photon density accumulates at the opposite end.

\textit{Stability phase diagram.--}
We address now the stability of the steady-state. 
Firstly, we define fluctuations $\bar{a}_j$ by the relation
$a_j = \alpha^{\rm ss}_j + \bar{a}_j$,
and consider the correlation matrix in the steady-state,
${M}_{j l} =   \langle \bar{a}^\dagger_j \bar{a}_l \rangle$, which evolves like
\begin{eqnarray}
\dot{M}_{j l}  = \sum_{j'} H^*_{j j'} M_{j' l} + 
\sum_{l'} H_{l l'} M_{j l'} + \ 2 \Gamma^{(p)}_{l j} .
\label{2point}
\end{eqnarray}
Together with Eq. \eqref{coherences}, Eq. \eqref{2point} provides us with a complete characterization of the steady-state.
$M_{j l}$ converges to a non-infinite value provided that $\Re\left(\lambda_n\right) < 0$.
Whereas the steady-state of the coherences can be analyzed by means of the s.v.d, see Eq. \eqref{steady.state}, fluctuations are directly governed by eigenvalues of the non-Hermitian matrix $H$.
%
%

\begin{figure}[h!]
  \includegraphics[width=0.5\textwidth]{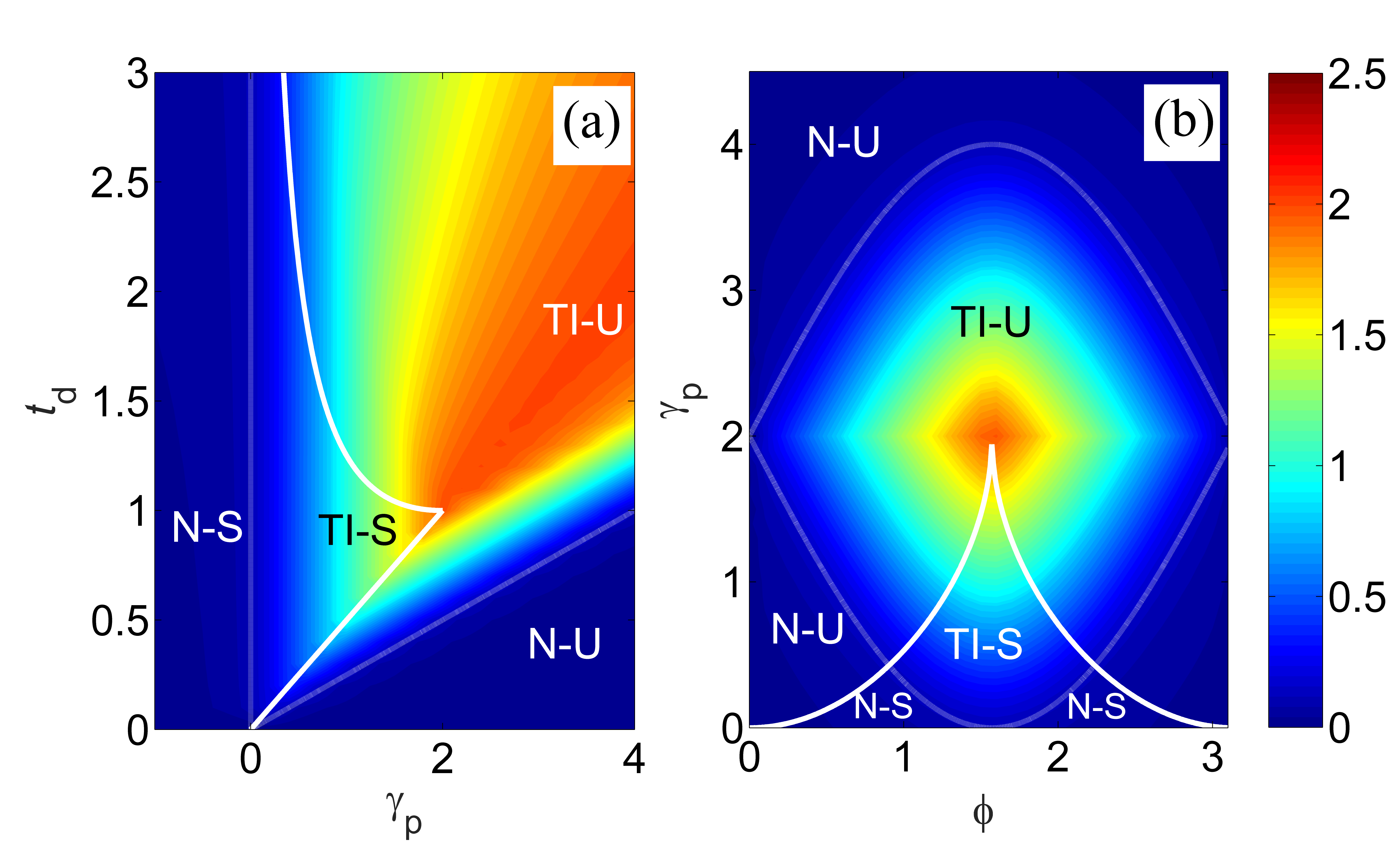}
  \caption{Singular value gap, $\Delta_s = s_{N-1} - s_N$, for the model defined by Eq. \eqref{1D.couplings},  $N = 100$. 
  Dashed lines separate Topological Insulator (TI) and Normal (N) phases. The intersection, TI-S, is the region of stable non-trivial topological phases. Continuous lines separate stable (S) and unstable (U) phases.
  (a) $\phi = \pi/2$, $t_\cc = 1$ and different values of $t_\dd$ and $\gamma_\pp$. 
  (b) $t_\dd = t_\cc = 1$ and different values of $\phi$ and $\gamma_\pp$.
} 
  \label{fig:fig2} 
\end{figure}

Our one-dimensional example \eqref{1D.couplings} can be exactly diagonalized \cite{noschese13}  
and we can seek topologically non-trivial stable regimes.
The eigenvalues of $H$  with periodic boundary conditions are  
$\lambda(k) = \gamma_\pp - 2 t_\dd + 2 t_\dd \cos(k) + 2 i t_\cc \cos(k + \phi)$. 
Condition $\Re (\lambda(k)) < 0$ requires that $\gamma_\pp < 0 $, which is not compatible with the existence of a non-trivial topological phase. However the situation radically changes when we consider open boundary conditions. Here an exact solution is also available, leading to a set of $n = 1, \dots, N$ eigenvalues of $H$,
\begin{eqnarray}
\lambda_n &=& \gamma_\pp - 2 t_\dd 
\nonumber \\
&+& 2 \sqrt{(i t_\cc e^{i \phi} \! - \! t_\dd)(i t_\cc e^{-i \phi} \! -  \! t_\dd)} 
\cos \left( \! \frac{n \pi}{N \! + \! 1} \! \right) .
\end{eqnarray}
%
Assume that $t_\cc = t_\dd$, then stable solutions exist if
\begin{equation}
\gamma_\pp < 2 t_\dd \left( 1 - \sqrt{\cos \phi} \right) .
\label{eq:condition_stablity}
\end{equation}
An overlapping region between the stable and non-trivial topological regimes defined by both conditions Eq. 
\eqref{eq:condition_stablity} and \eqref{eq:condition_topology} can be found as long as
$
|\cos (\phi)| < (-1 + \sqrt{5})/2 \approx 0.62 \ .
$
Our model thus requires a threshold minimum value of $\phi$ for stable non-trivial topological phases to exist. Numerical calculations in a wide range of parameters confirm the existence of stable topologically nontrivial phases in the steady-state phase diagram, see Fig. \ref{fig:fig2}.

\begin{figure}[h!]
	\includegraphics[width=0.45\textwidth]{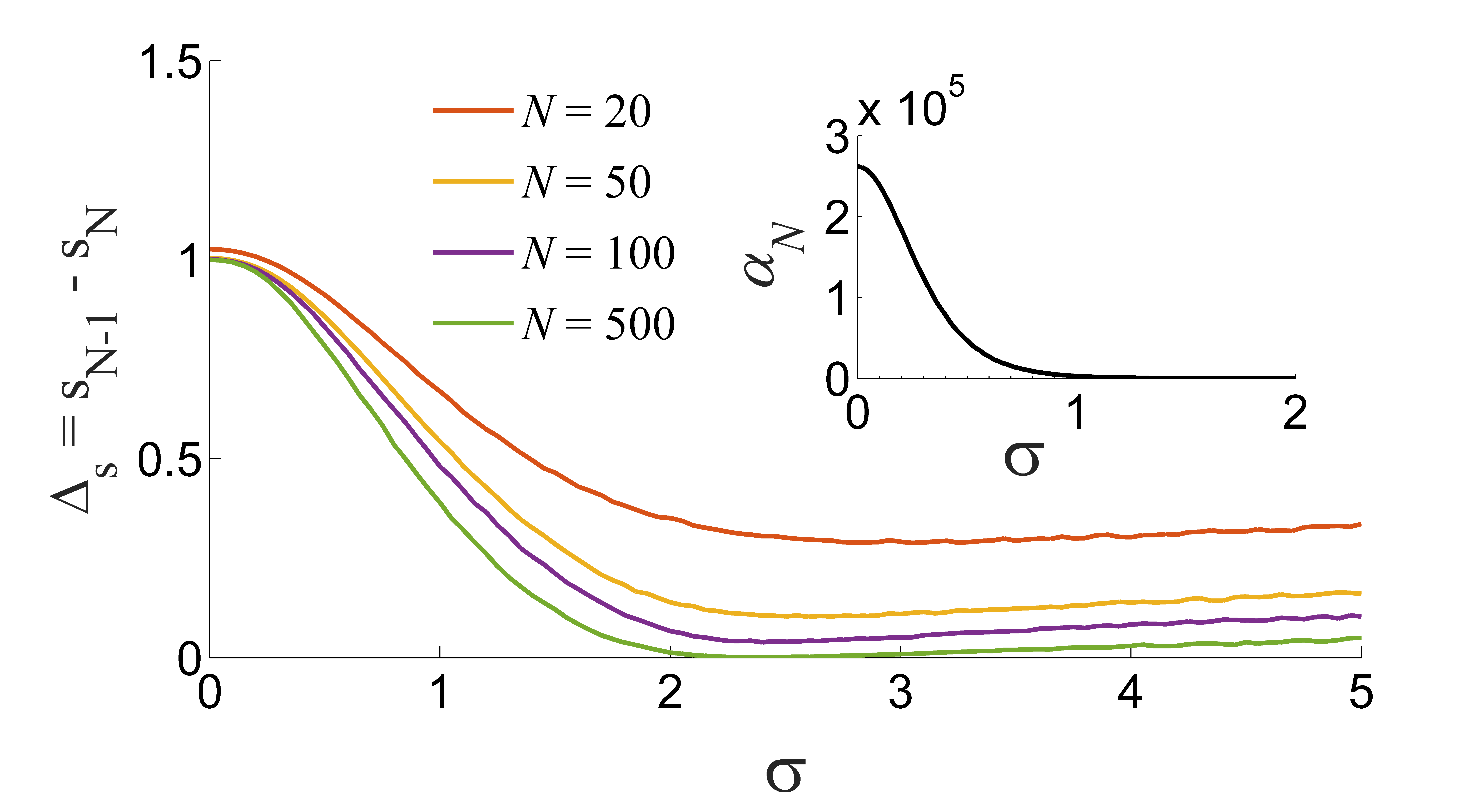}
	\caption{Gap between the second and lowest singular value of $H$ given by Eqs. \eqref{1D.couplings} with $t_\cc = t_\dd = \gamma_\pp = 1$, as a function of disorder strength, averaged over $10^5/N$ realizations of disorder.
	Inset: Field amplitude at site $j = N$ induced by a coherent drive at $j = 1$ with $\epsilon_1 = 1$ in a chain with $N = 100$.}
	\label{fig:fig3} 
\end{figure}

\textit{Topological protection.--} 
We check the robustness of topological amplification by adding  a diagonal disorder term to the Hermitian coupling matrix $G$, 
$\delta G_{j,l} = \delta \omega_{j}  \delta_{j,l}$, 
where $\delta \omega_j$ are Gaussian random variables with standard deviation $\sigma$. 
We calculate the gap $\Delta_{s} = s_{N-1} - s_N$ for increasing number of lattices sites  
(Fig. \ref{fig:fig3}), and find that the topological phase is robust for a wide range of $\sigma$.

\textit{Physical implementations.--} 
Our ideas can be implemented by using schemes for non-reciprocal transport and directional amplification \cite{Devoret11,Devoret13,Devoret15,Metelmann15,Anderson16,Owens18,Lee16,Nunnenkamp18}.
Complex photon tunneling can be induced by using periodic modulations in circuit QED \cite{Quijandria13prl,Peropadre13,navarrete14prl,Metelmann17,Quijandria18} or even vibronic lattices \cite{Bermudez11prl,Bermudez12}.
This approach allows for smaller system sizes than, e.g., the use of magnetic fields.

We propose an implementation motivated by the superconducting circuit setup presented in \cite{roushan17}. We describe below the main ideas - technical details can be found in  the Appendix..

\begin{figure}[h!]
	\includegraphics[width=0.5\textwidth]{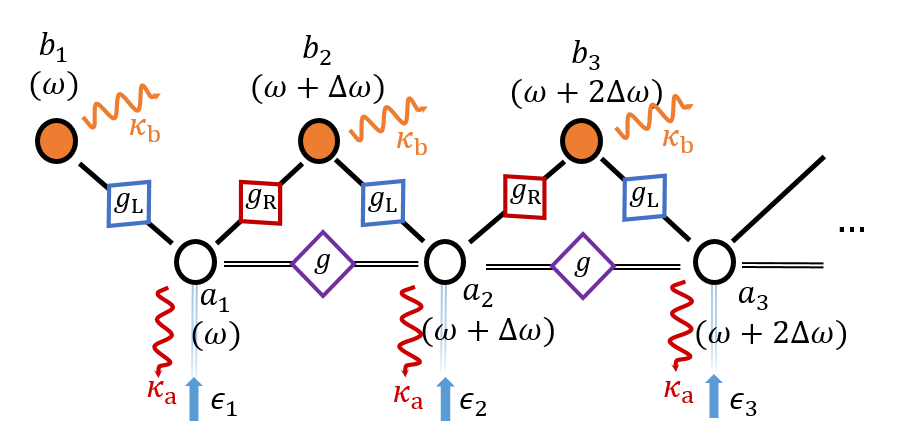}
	\caption{Scheme for a physical implementation of the 1D topological amplifier \eqref{1D.couplings}. White circles: main local photonic modes. Orange circles:  auxiliary fast decaying modes for reservoir engineering.}
	\label{fig:fig4} 
\end{figure}

The main system is a chain of $N$ cavities (mode operators $a_j$) coupled to an auxiliary chain (modes $b_j$). 
Cavities are arranged in the ladder configuration of Fig. \ref{fig:fig4}. 
The non-interacting cavity system is described by
\begin{equation}
H_0 = \sum_{j=1}^N \omega_j  a^{\dagger}_j a_j + 
 \sum_{j=1}^{N+1} \omega_j  b^{\dagger}_j b_j.
\end{equation}
Nearby cavity frequencies are separated by $\Delta \omega$,
such that $\omega_j = \omega + \Delta \omega (j-1))$. The frequency gradient is used to break time-reversal symmetry. Both main and auxiliary cavities are subjected to photon leakage with rates $\kappa_{\rm a}$ and $\kappa_{\rm b}$, respectively.

Complex coherent couplings are induced by periodic modulations,
\begin{equation}
H_{\rm c}(t) = 
g(t) \sum_{j=1}^N
(a_j + a_j^\dagger ) ( a_{j+1} + a_{j+1}^\dagger ),
\end{equation} 
with $g(t) = g_0   \cos(\Delta \omega t + \phi_{\rm d})$. In the interaction picture with respect to $H_0$ and in a r.w.a. valid if
$\Delta \omega \gg g_0$, we obtain the photon tunneling terms of Eq. \eqref{1D.couplings} with $t_\cc = g_0/2$ and $\phi = \phi_{\rm d}$.

The auxiliary cavities will provide us with collective incoherent pumping. We consider the couplings
\begin{eqnarray}
\hspace{-1cm} H_{\rm aux}(t) 
&=&
\sum_{j=1}^N g_{{\rm R},j}(t) ( a_j + a_j^\dagger ) ( b_{j+1} + b_{j+1}^\dagger ) 
\nonumber \\
&+&
\sum_{j=1}^N g_{{\rm L},j}(t) ( a_j + a_j^\dagger ) ( b_j + b_j^\dagger ) ,
\end{eqnarray}
with periodic modulations 
$g_{{\rm R},j}(t) =  \bar{g}_{0} \cos((\omega_j + \omega_{j+1})t)$, 
$g_{{\rm L},j}(t) =  \bar{g}_{0} \cos(2 \omega_j t).$
In the interaction picture with respect to $H_0$, and in a rotating wave approximation (r.w.a.) valid if
$\bar{g}_0 \ll \omega_j$ we get
\begin{equation}
H'_{\rm aux} \underrel{\rm r.w.a.}{=}
\frac{\bar{g}_{0}}{2} 
\sum_{j=1}^N \left(a_j (b_j + b_{j+1} \right) + \rm{H.c.}  ).
\end{equation}
Consider now that auxiliary $b$-modes are very fast decaying ($\bar{g}_0 \ll \kappa_{\rm b}$), such that we can adiabatically eliminate them  \cite{cirac92} and get
dissipative couplings 
$\Gamma_{j l} = (2 t_\dd-\kappa_{\rm a}) \delta_{j l} + t_\dd \delta_{l,j+1} + t_\dd \delta_{l,j-1}$, with 
 $t_{\rm d} = (\bar{g}_0)^2/(4 \kappa_{\rm b})$ (see Appendix for details). We thus obtain a collective incoherent pumping induced by virtual photons being emitted into the auxiliary cavities. Finally, we can obtain the dissipative coupling matrix in the form of Eq. \eqref{1D.couplings}
by defining the net pumping rate $\gamma_{\rm p} = 2 t_\dd - \kappa_{\rm a}$, such that $\gamma_{\rm p} = 0$ is precisely the value at which gain and loss are balanced.

Topologically phases can be detected, for example by adding a coherent drive at one edge and measuring the coherences, $\langle a_j \rangle$, in the steady state by homodyne detection. Exponential amplification would signal the existence of a singular edge state.

\textit{Conclusions and Outlook.--}
This work presents a connection between directional amplification and topological insulator theory. Our approach leads to a classification of topological phases of non-Hermitian matrices that is directly connected to applications of photonic lattices as amplifiers. 
We have presented an example that could be implemented with circuit QED setups.
In the future we aim to investigate many-body effects \cite{fitzpatrick17}, lasing phases \cite{Fernandez18}, disorder and/or long range interactions and couplings \cite{Nevado17}.

Funded by the People Programme of the EU’s Seventh Framework Programme, Ref. PCIG14-GA-2013-630955.

	\appendix

	\section{Physical implementation of a topological amplifier: Floquet and reservoir engineering}
	In this appendix section we propose in more detail a particular implementation of our one-dimensional model in Eq. \eqref{1D.couplings}. 
	For ease of reading, in this document we will reproduce some of the equations from the main article.
	
	Our proposal relies on the use of periodic drivings and Floquet engineering. 
	It could be implemented directly in the microwave regime with a circuit QED experimental setup like the one used in Ref. \cite{roushan17} (Martinis' group). However, the same or very similar ideas could be implemented in other circuit QED setups \cite{Quijandria13prl,Peropadre13,navarrete14prl,Metelmann17,Quijandria18} or even vibronic lattices \cite{Bermudez11prl,Bermudez12}. 
	
	In the scheme presented here periodic drivings induce complex photon tunneling terms without the need of using magnetic fields. The latter typically impose limitations in the minimum size of the device. Thus, avoiding magnetic fields is advantageous in implementations which may aim to observe quantum effects or be used as quantum detectors. 
	In addition to induce complex photon tunneling, we also need to control dissipative couplings. 
	As we show below, this can be achieved by introducing a set of auxiliary modes that are used for reservoir 
	engineering. 
	
	Our scheme employs two arrays of photonic cavities (see Fig. \ref{fig:fig4}). For one hand, we use $N$ main cavities, hosting local modes with annihilation operators, $a_j$, to directly implement the dissipative topological phases.  On the other hand we introduce an array of $N+1$ auxiliary photonic cavities with local modes, $b_j$, which we use to induce dissipative couplings amongst the main cavities, 
	see Fig. \ref{fig:fig4}. 
	We assume that both the main and auxiliary cavities have frequencies that vary linearly with the cavity position,
	\begin{eqnarray}
	H_0 = \sum_{j=1}^N \omega_j a^{\dagger}_j a_j 
	\! + \! \sum_{j=1}^{N+1} \omega_j  b^{\dagger}_j b_j , \ 
	\omega_j = \omega + (j\!-\!1) \Delta \omega.
	\end{eqnarray}
	Nearby cavity frequencies differ by a step, $\Delta \omega$, which will allow us to break time-reversal invariance and control the phase of the complex tunnelings. We also assume that both the main and the auxiliary cavities are subjected to photon leakage with rates $\kappa_{\rm a}$ and $\kappa_{\rm b}$, respectively. The Liouvillian, 
	${\cal L}_0$, describing uncoupled cavities reads
	\begin{eqnarray}
	{\cal L}_0 (\rho) = - i [H_0, \rho ]  + {\cal L}_{\kappa_{\rm a}}(\rho) 
	+  {\cal L}_{\kappa_{\rm b}}(\rho),
	\end{eqnarray}
	with
	\begin{eqnarray}
	{\cal L}_{\kappa_{\rm a}} (\rho) &=& \kappa_{\rm a} \sum_{j=1}^N 
	\left( 2 a_j \rho a_j^\dagger -  a_j^\dagger a_j \rho - \rho a_j^\dagger a_j  \right) \nonumber \\
	{\cal L}_{\kappa_{\rm b}} (\rho) &=& \kappa_{\rm a} \sum_{j=1}^{N+1} 
	\left( 2 b_j \rho b_j^\dagger -  b_j^\dagger b_j \rho - \rho b_j^\dagger b_j  \right) . 
	\end{eqnarray}
	Later on we will work in a limit in which auxiliary cavities are in a ``bad" cavity limit of $\kappa_{\rm b}$ being large compared to other energy scales in the system.
	
	%
	%

	We introduce two set of couplings: $H_{\rm c}(t)$ is a time-dependent coupling between the main $a$-cavities, whereas $H_{\rm aux}(t)$ is a coupling term between auxiliary $b$-cavities and the main $a$-cavities. We assume that both couplings are time-dependent. Below we show how a judicious choice of time-dependent couplings and decay rates leads directly to Eq. \eqref{1D.couplings}.
	
	\subsection{Complex photon tunneling terms}
	The coupling between cavities is subjected to a periodic modulation,

	\begin{equation}
	H_{\rm c}(t) = 
	g(t) \sum_{j=1}^{N-1} 
	\left(a_j + a_j^\dagger \right) \left( a_{j+1} + a_{j+1}^\dagger \right).
	\end{equation} 
	We assume a periodic modulation with a frequency that matches the frequency difference between nearby cavities and a finite phase $\phi$,
	\begin{equation}
	g(t) = g_0   \cos(\Delta \omega t + \phi) .
	\end{equation}
	It has been previously shown that such a choice of time-dependent couplings can be used to induce a complex photon tunneling term with phase $\phi$, see for example theoretical proposals 
	\cite{fang12,Peropadre13} and the circuit QED experimental result published in \cite{roushan17}.

	We select the resonant terms by direct inspection of the coherent coupling expressed in the interaction picture with respect to $H_0$,
	\begin{eqnarray}
	H'_{\rm c}(t) &=& 
	g_{\rm 0} \cos(\Delta \omega t + \phi) 
	\sum_j \left(a^\dagger_j a_{j+1} e^{-i \Delta \omega t}  + {\rm H.c.} \right)  \nonumber \\
	&\underrel{\rm r.w.a.}{=}&
	\frac{g_{\rm 0}}{2} \sum_j \left(a^\dagger_j a_{j+1} e^{i \phi}  + {\rm H.c.} \right),
	\end{eqnarray} 
	where we have assumed the validity of the rotating wave approximation (r.w.a.) and neglected terms evolving like $2 \Delta \omega$, an assumption that holds in the limit 
	\begin{equation}
	\Delta \omega \gg g_0, \kappa .
	\end{equation}
	After the r.w.a. we thus arrive at the complex photon tunneling term in Eq. \eqref{1D.couplings} with phase $\phi$ determined by the periodic modulation of the coupling and $t_\cc = g_0/2$.

	\subsection{Dissipative coupling}
	We also need to introduce a periodic modulation of the coupling between the main and the auxiliary cavities. 
	However, contrary to the previous case, we aim to select now counter-rotating terms of the form $a_j b_j$,  $a^\dagger_j b^\dagger_j$. 
	Our goal here is to use the auxiliary cavity array both to inducing the dissipative couplings between nearest neighbor cavities, $t_{\rm d}$, and at the same time to inducing the incoherent pump term, $\gamma_{\rm p}$ in Eq. \eqref{1D.couplings}. This is the most economic choice to implement all the required dissipative elements with a minimum need of auxiliary subsystems for reservoir engineering.
	
	To meet our goals we assume the configuration of coupled triangles shown in Fig. \ref{fig:fig4}, which leads to $a$-$b$ couplings of the form,
	\begin{eqnarray}
	\hspace{-1cm} H_{\rm aux}(t) 
	&=&
	\sum_{j=1}^N g_{{\rm R},j}(t) \left( a_j + a_j^\dagger \right) \left( b_{j+1} + b_{j+1}^\dagger \right) 
	\nonumber \\
	&+&
	\sum_{j=1}^N g_{{\rm L},j}(t) \left( a_j + a_j^\dagger \right) \left( b_j + b_j^\dagger \right) .
	\end{eqnarray}
	We choose a periodic modulation of the couplings with the same amplitude, $\bar{g}_0$, and frequencies tuned such that non-conserving photon couplings are resonant,
	\begin{eqnarray}
	g_{{\rm R},j}(t) &=& \bar{g}_{0} \cos((\omega_j + \omega_{j+1})t) , \nonumber \\
	g_{{\rm L},j}(t) &=& \bar{g}_{0} \cos(2 \omega_j t).
	\end{eqnarray}
	Again we can express those coupling terms in the interaction picture with respect to $H_0$ and neglect fast rotating terms, such that we finally get,
	\begin{eqnarray}
	H'_{\rm aux} &\underrel{\rm r.w.a.}{=}&
	\sum_{j = 1}^N
	\frac{\bar{g}_{0}}{2} 
	\left(a_j \left(b_j + b_{j+1} \right) + \rm{H.c.}  \right).
	\end{eqnarray} 
	This approximation is valid in the limit $\bar{g}_0 \ll \omega_j$.
	In the interaction picture with respect to $H_0$ and under the r.w.a., the dynamics of the joint main and auxiliary systems is thus described by the following Liouvillian,
	%
	\begin{eqnarray}
	{\cal L}(\rho) &=& -i [H_{\rm c}' + H'_{\rm aux}, \rho] 
	+ {\cal L}_{\kappa_a}(\rho) + {\cal L}_{\kappa_b}(\rho) .
	\label{before.adiab}
	\end{eqnarray}

	For the next step of our derivation we neeed to assume that auxiliary cavities decay very fast compared to the couplings to the main cavities, 
	\begin{equation}
	\kappa_{\rm b} \gg \bar{g}_0/2.
	\end{equation}
	Under this condition we can adiabatically eliminate the auxiliary, $b$-modes, which can be assumed to be in the vacuum state. Applying perturbation theory in the coupling term, $H_{\rm aux}$, we can find an approximate master equation for the reduced density matrix of the main, $a$-modes, 
	$\rho_{\rm a} = {\rm Tr}_{\rm b}(\rho)$, which reads
	\begin{eqnarray}
	{\cal L} (\rho_{\rm a}) = 
	-i [H_{\rm c}', \rho_{\rm a}] 
	&+& {\cal L}_{\kappa_a}(\rho_{\rm a})  \nonumber \\
	&& \hspace{-1cm}  + 
	\frac{(\bar{g}_0)^2}{4 \kappa} 
	\sum_{j = 1}^{N-1} \left( 
	2 (a^\dagger_j + a^\dagger_{j+1}) \rho_{\rm a} (a_j + a_{j+1}) \right.
	\nonumber \\  
	\hspace{1cm} &-& (a_j + a_{j+1}) (a^\dagger_j + a^\dagger_{j+1}) \rho_{\rm a} 
	\nonumber \\  
	\hspace{1cm} &-& \left.  \rho_{\rm a} (a_j + a_{j+1}) (a^\dagger_j + a^\dagger_{j+1}) \right).
	\label{adiab}
	\end{eqnarray}
	Eq. \eqref{adiab} is a very intuitive result describing the coupling of the main cavities by the virtual emission/absorption of a fast decaying auxiliary $b$-photon. 
	To prove this result we have to use perturbation theory in Liouvillian space as carried out, e.g., in the theory of laser cooling \cite{cirac92}. A formal derivation is provided in the next section. For the moment, let us analyze the couplings obtained in Eq. \eqref{adiab}. Let us first define
	\begin{equation}
	t_\dd = \frac{(\bar{g}_0)^2}{4 \kappa_{\rm a}} .
	\end{equation}
	If we use the parametrization of Eq. \eqref{photonic.lattice} we obtain the matrices
	\begin{eqnarray}
	\Gamma^{(\rm p)}_{jl} &=& 2 t_\dd \delta_{j,l} + t_\dd \delta_{j,l+1} + t_\dd \delta_{j,l-1} , 
	\nonumber \\
	\Gamma^{(\rm d)}_{jl} &=&  k_{\rm a} \delta_{j,l} .
	\end{eqnarray}
	To obtain the matrix $\Gamma_{jl}$ in \eqref{1D.couplings} we define 
	$\gamma_{\rm p} = 4 t_{\rm d} - \kappa_{\rm a}$, such that
	\begin{equation}
	\Gamma_{jl} = \Gamma^{(\rm p)}_{jl} \! - \! \Gamma^{(\rm d)}_{jl} = (\gamma_{\rm p} \! - \! 2 t_\dd) \delta_{j,l} + t_\dd \delta_{j,l+1} + t_\dd \delta_{j,l-1} .
	\label{good.param}
	\end{equation}
	Dissipative terms are thus parametrized by $t_{\rm d}$ and $\gamma_{\rm p}$, rather than by 
	$t_{\rm d}$ and $\kappa_{\rm a}$. 
	This parametrization makes the discussion of different topological phases more transparetn. Specifically $\gamma_{\rm p}$ represents the net incoherent photon pumping in the lattice. Let us consider the eigenstates of $\Gamma_{jl}$ in the case of periodic boundary conditions,
	\begin{equation}
	\Gamma(k) = (\gamma_{\rm p} - 2 t_{\rm d}) + 2 t_{\rm d} \cos(k).
	\end{equation}
	If $\gamma_{\rm p} > 0$, we find values $\Gamma(k) > 0$, indicating a net positive photon pumping. The case $\kappa_{\rm a} = 4 t_{\rm d} \to \gamma_{\rm p} = 0$, actually corresponds to a situation in which the net photon gain induced by the auxiliary cavities is exactly canceled by the photon loss in the main cavities. 
	
	\subsection{Coherent driving of the main cavities}
	We have shown that the photonic lattice described by Eq. \eqref{adiab} yields the same non-Hermitian coupling matrix, $H$, defined in our 1D example, Eqs. \eqref{1D.couplings}. However, we need to add the coherent drivings that will be subjected to the topological amplification mechanism presented in the main text. Assume a set of coherent fields with frequencies $\Omega_j$ and complex amplitudes such that the coupling with the main cavities has the form
	\begin{equation}
	H_{\epsilon}(t) = \sum_{j} \left( \epsilon_j e^{-i \Omega_j t} + \epsilon^*_j e^{i \Omega_j t} \right) \left( a_j + a_j^\dagger\right).
	\end{equation}
	In the previous subsections we have been working in the interaction picture with respect to $H_0$. In that picture and after a r.w.a. we get
	\begin{equation}
	H_{\epsilon}'(t) \underrel{\rm r.w.a.}{=}
	\sum_{j} \left( \epsilon_j a^\dagger_j e^{-i (\Omega_j-\omega_j) t}
	+ \epsilon^*_j a_j e^{i (\Omega_j -\omega_j)t} \right) .
	\end{equation}
	To recover the case that we have been analyzing in the main text (that is, resonant coherent fields) we would need driving frequencies on resonance with the linearly varying cavity frequencies 
	($\Omega_j = \omega + (\Delta \omega) (j -1)$).
	
	In applications where a drive with a single frequency, $\omega$ needs to be amplified, one could illuminate only a single cavity, for example $j = 1$, such that $\epsilon_j = \epsilon \delta_{j,1}$ and $\Omega_1 = \omega$. Choosing now a localized singular edge mode would allow to induce the topological amplification effect in the resonant limit studied in the text, without the need of using a multi-frequency coherent drive.
	\subsection{Circuit QED implementation}
	As mentioned above the scheme presented in the last subsections can actually be applied to a variety of implementations. The specific configuration chosen here has the advantage that it can be directly realized in the experimental setup developed in Ref. \cite{roushan17}. 
	That reference shows the operation of the basic building block of our scheme, namely, one of the triangular cavity configurations, together with periodic modulations of the couplings, leading to the implementation of a synthetic gauge field.
	
	The required energy hierarchy for our proposal is
	\begin{eqnarray}
	\bar{g}_0 &\ll& \kappa_{\rm b} \ll \omega \nonumber \\
	\kappa_{\rm a}, g_0 &\ll& \Delta \omega \ll \Omega .
	\end{eqnarray}
	In \cite{roushan17}, couplings are $\{\bar{g}_0$, $g_0\} = 4$ MHz,  $\Delta \omega =$ 35 MHz, and  $\omega = 5.8$ GHz. Photon loss is very slow in the experiment, so that 
	$\{\kappa_{\rm a}$, $\kappa_{\rm b} \} = 0.1$ MHz. The photon decay rates are, however, easily tuned by fabrication, for example, by enhancing the coupling rate of the superconducting cavities to the transmission lines that are used for measurement or excitation. Values of $\kappa_{\rm a} \approx 4$ MHz and $\kappa_{\rm b} \approx 40$ MHz would be enough to bring the system in the desired parameter regime. 
	
	Finally in Ref. \cite{roushan17}, the experiment is carried out in a regime in which superconducting cavities have a strong non-linear term. The latter can be suppressed in circuit QED setups to bring the system into the linear regime considered in this work. In any case, the inclusion on non-linearities leading to photon-photon interaction is an interesting feature that would lead to richer physics related to interacting topological phases.

	\section{Adiabatic elimmination of auxiliary subsystems.}
	We present now the necessary steps to eliminate adiabatically the auxiliary cavities hosting fast decaying modes,  $b_j$. Our starting point is Eq. \eqref{before.adiab}. We follow here the formalism introduced by Cirac {\it et al.} for the description of laser cooling in \cite{cirac92}.
	
	The basic idea is to carry out perturbation theory in Liouville space. We focus on  the following master equation,
	\begin{equation}
	\dot{\rho} = {\cal L}(\rho) = {\cal L}_{\kappa_{\rm b}}(\rho) + {\cal L}_{\rm I}(\rho) ,
	\end{equation}
	which includes the fast photon leakage out of the auxiliary cavities, plus the interaction of the auxiliary and the main cavities, 
	${\mathcal{L}}_{\rm I}(\rho) = - i [H_{\rm aux}, \rho]$.
	We are considering the limit $\kappa_{\rm b} \gg \bar{g}_0$, such that ${\cal L}_{\rm I}$ can be considered a small perturbation to the fast auxiliary mode decay.
	
	We define the projection operator in Liouville space, 
	\begin{equation}
	{\cal P} \rho = \rho_{\rm a}  \otimes |0\rangle_{\rm b} \langle 0|,
	\end{equation}
	which projects a density matrix in a product state of the reduced density matrix of the main modes times the steady state of ${\cal L}_{\kappa_{\rm b}}$. Now we proceed by finding a perturbative equation for the dynamics of ${\cal P}\rho$. 
	For this we define first the complementary projection operator
	${\cal Q} = \mathbb{1} - {\cal P}$, and write
	\begin{eqnarray}
	{\cal P}\dot{\rho} &=& {\cal P} {\cal L}_{\rm I}  {\cal Q} \rho ,
	\nonumber \\
	{\cal Q}\dot{\rho} &=& {\cal Q} {\cal L}_{\kappa_{\rm b}}  {\cal Q} \rho + 
	{\cal Q} {\cal L}_{\rm I}  {\cal P} \rho + 
	{\cal Q} {\cal L}_{\rm I}  {\cal Q} \rho,
	\end{eqnarray}
	where we have used that ${\cal L}_{\kappa_{\rm b}}{\cal P} \rho = 0$ and 
	${\cal P}{\cal L}_{\rm I}{\cal P} = 0$. To second order in the perturbative term, we can write
	\begin{equation}
	{\cal P} \dot{\rho} = 
	- \int^t_0 d\tau 
	{{\cal P} \cal L}_{\rm I} {\cal Q} 
	e^{{\cal L}_{\kappa_{\rm b}} \tau} {\cal Q} {\cal L}_{\rm I} {\cal P} \rho .
	\end{equation}
	We can use this equation to obtain a master equation for the reduced density matrix,
	\begin{equation}
	\dot{\rho}_{\rm a} =\int^t_0 d\tau {\rm Tr} \left(
	[H_{\rm aux}, e^{{\cal L}_{\kappa_{\rm b}} \tau}
	\left( [H_{\rm aux}, \rho_{\rm a} \otimes | 0 \rangle_{\rm b} \langle0|]\right) ]  \right)
	\label{adiab.l}
	\end{equation}
	Eq. \eqref{adiab.l}, together with the quantum regression theorem allows us to write the different terms within the time integral in terms of two-time correlation functions of the auxiliary modes under the dynamics of ${\cal L}_{\kappa_{\rm b}}$,
	\begin{equation}
	\langle b_j(\tau) b^\dagger_l (0)\rangle = 
	\delta_{j,l} e^{-\kappa_{\rm b} \tau}.
	\end{equation}
	Using the above result and after some rather straightforward algebra we arrive at the equation for the reduced density matrix, Eq. \eqref{adiab}.

	\bibliographystyle{apsrev4-1}
	\bibliography{biblio}

\begin{thebibliography}{63}%
\makeatletter
\providecommand \@ifxundefined [1]{%
 \@ifx{#1\undefined}
}%
\providecommand \@ifnum [1]{%
 \ifnum #1\expandafter \@firstoftwo
 \else \expandafter \@secondoftwo
 \fi
}%
\providecommand \@ifx [1]{%
 \ifx #1\expandafter \@firstoftwo
 \else \expandafter \@secondoftwo
 \fi
}%
\providecommand \natexlab [1]{#1}%
\providecommand \enquote  [1]{``#1''}%
\providecommand \bibnamefont  [1]{#1}%
\providecommand \bibfnamefont [1]{#1}%
\providecommand \citenamefont [1]{#1}%
\providecommand \href@noop [0]{\@secondoftwo}%
\providecommand \href [0]{\begingroup \@sanitize@url \@href}%
\providecommand \@href[1]{\@@startlink{#1}\@@href}%
\providecommand \@@href[1]{\endgroup#1\@@endlink}%
\providecommand \@sanitize@url [0]{\catcode `\\12\catcode `\$12\catcode
  `\&12\catcode `\#12\catcode `\^12\catcode `\_12\catcode `\%12\relax}%
\providecommand \@@startlink[1]{}%
\providecommand \@@endlink[0]{}%
\providecommand \url  [0]{\begingroup\@sanitize@url \@url }%
\providecommand \@url [1]{\endgroup\@href {#1}{\urlprefix }}%
\providecommand \urlprefix  [0]{URL }%
\providecommand \Eprint [0]{\href }%
\providecommand \doibase [0]{http://dx.doi.org/}%
\providecommand \selectlanguage [0]{\@gobble}%
\providecommand \bibinfo  [0]{\@secondoftwo}%
\providecommand \bibfield  [0]{\@secondoftwo}%
\providecommand \translation [1]{[#1]}%
\providecommand \BibitemOpen [0]{}%
\providecommand \bibitemStop [0]{}%
\providecommand \bibitemNoStop [0]{.\EOS\space}%
\providecommand \EOS [0]{\spacefactor3000\relax}%
\providecommand \BibitemShut  [1]{\csname bibitem#1\endcsname}%
\let\auto@bib@innerbib\@empty
\bibitem [{\citenamefont {{Ozawa}}\ \emph {et~al.}(2018)\citenamefont
  {{Ozawa}}, \citenamefont {{Price}}, \citenamefont {{Amo}}, \citenamefont
  {{Goldman}}, \citenamefont {{Hafezi}}, \citenamefont {{Lu}}, \citenamefont
  {{Rechtsman}}, \citenamefont {{Schuster}}, \citenamefont {{Simon}},
  \citenamefont {{Zilberberg}},\ and\ \citenamefont {{Carusotto}}}]{ozawa18}%
  \BibitemOpen
  \bibfield  {author} {\bibinfo {author} {\bibfnamefont {T.}~\bibnamefont
  {{Ozawa}}}, \bibinfo {author} {\bibfnamefont {H.~M.}\ \bibnamefont
  {{Price}}}, \bibinfo {author} {\bibfnamefont {A.}~\bibnamefont {{Amo}}},
  \bibinfo {author} {\bibfnamefont {N.}~\bibnamefont {{Goldman}}}, \bibinfo
  {author} {\bibfnamefont {M.}~\bibnamefont {{Hafezi}}}, \bibinfo {author}
  {\bibfnamefont {L.}~\bibnamefont {{Lu}}}, \bibinfo {author} {\bibfnamefont
  {M.}~\bibnamefont {{Rechtsman}}}, \bibinfo {author} {\bibfnamefont
  {D.}~\bibnamefont {{Schuster}}}, \bibinfo {author} {\bibfnamefont
  {J.}~\bibnamefont {{Simon}}}, \bibinfo {author} {\bibfnamefont
  {O.}~\bibnamefont {{Zilberberg}}}, \ and\ \bibinfo {author} {\bibfnamefont
  {I.}~\bibnamefont {{Carusotto}}},\ }\href@noop {} {\bibfield  {journal}
  {\bibinfo  {journal} {ArXiv e-prints}\ } (\bibinfo {year} {2018})},\ \Eprint
  {http://arxiv.org/abs/1802.04173} {arXiv:1802.04173 [physics.optics]}
  \BibitemShut {NoStop}%
\bibitem [{\citenamefont {Schnyder}\ \emph {et~al.}(2008)\citenamefont
  {Schnyder}, \citenamefont {Ryu}, \citenamefont {Furusaki},\ and\
  \citenamefont {Ludwig}}]{Schnyder08}%
  \BibitemOpen
  \bibfield  {author} {\bibinfo {author} {\bibfnamefont {A.~P.}\ \bibnamefont
  {Schnyder}}, \bibinfo {author} {\bibfnamefont {S.}~\bibnamefont {Ryu}},
  \bibinfo {author} {\bibfnamefont {A.}~\bibnamefont {Furusaki}}, \ and\
  \bibinfo {author} {\bibfnamefont {A.~W.~W.}\ \bibnamefont {Ludwig}},\ }\href
  {\doibase 10.1103/PhysRevB.78.195125} {\bibfield  {journal} {\bibinfo
  {journal} {Phys. Rev. B}\ }\textbf {\bibinfo {volume} {78}},\ \bibinfo
  {pages} {195125} (\bibinfo {year} {2008})}\BibitemShut {NoStop}%
\bibitem [{\citenamefont {Ryu}\ \emph {et~al.}(2010)\citenamefont {Ryu},
  \citenamefont {Schnyder}, \citenamefont {Furusaki},\ and\ \citenamefont
  {Ludwig}}]{ryu10}%
  \BibitemOpen
  \bibfield  {author} {\bibinfo {author} {\bibfnamefont {S.}~\bibnamefont
  {Ryu}}, \bibinfo {author} {\bibfnamefont {A.~P.}\ \bibnamefont {Schnyder}},
  \bibinfo {author} {\bibfnamefont {A.}~\bibnamefont {Furusaki}}, \ and\
  \bibinfo {author} {\bibfnamefont {A.~W.~W.}\ \bibnamefont {Ludwig}},\ }\href
  {http://stacks.iop.org/1367-2630/12/i=6/a=065010} {\bibfield  {journal}
  {\bibinfo  {journal} {New J. Phys.}\ }\textbf {\bibinfo {volume} {12}},\
  \bibinfo {pages} {065010} (\bibinfo {year} {2010})}\BibitemShut {NoStop}%
\bibitem [{\citenamefont {Bansil}\ \emph {et~al.}(2016)\citenamefont {Bansil},
  \citenamefont {Lin},\ and\ \citenamefont {Das}}]{Bansil16}%
  \BibitemOpen
  \bibfield  {author} {\bibinfo {author} {\bibfnamefont {A.}~\bibnamefont
  {Bansil}}, \bibinfo {author} {\bibfnamefont {H.}~\bibnamefont {Lin}}, \ and\
  \bibinfo {author} {\bibfnamefont {T.}~\bibnamefont {Das}},\ }\href {\doibase
  10.1103/RevModPhys.88.021004} {\bibfield  {journal} {\bibinfo  {journal}
  {Rev. Mod. Phys.}\ }\textbf {\bibinfo {volume} {88}},\ \bibinfo {pages}
  {021004} (\bibinfo {year} {2016})}\BibitemShut {NoStop}%
\bibitem [{\citenamefont {Chiu}\ \emph {et~al.}(2016)\citenamefont {Chiu},
  \citenamefont {Teo}, \citenamefont {Schnyder},\ and\ \citenamefont
  {Ryu}}]{Chiu16}%
  \BibitemOpen
  \bibfield  {author} {\bibinfo {author} {\bibfnamefont {C.-K.}\ \bibnamefont
  {Chiu}}, \bibinfo {author} {\bibfnamefont {J.~C.~Y.}\ \bibnamefont {Teo}},
  \bibinfo {author} {\bibfnamefont {A.~P.}\ \bibnamefont {Schnyder}}, \ and\
  \bibinfo {author} {\bibfnamefont {S.}~\bibnamefont {Ryu}},\ }\href {\doibase
  10.1103/RevModPhys.88.035005} {\bibfield  {journal} {\bibinfo  {journal}
  {Rev. Mod. Phys.}\ }\textbf {\bibinfo {volume} {88}},\ \bibinfo {pages}
  {035005} (\bibinfo {year} {2016})}\BibitemShut {NoStop}%
\bibitem [{\citenamefont {Haldane}\ and\ \citenamefont
  {Raghu}(2008)}]{Haldane08}%
  \BibitemOpen
  \bibfield  {author} {\bibinfo {author} {\bibfnamefont {F.~D.~M.}\
  \bibnamefont {Haldane}}\ and\ \bibinfo {author} {\bibfnamefont
  {S.}~\bibnamefont {Raghu}},\ }\href {\doibase 10.1103/PhysRevLett.100.013904}
  {\bibfield  {journal} {\bibinfo  {journal} {Phys. Rev. Lett.}\ }\textbf
  {\bibinfo {volume} {100}},\ \bibinfo {pages} {013904} (\bibinfo {year}
  {2008})}\BibitemShut {NoStop}%
\bibitem [{\citenamefont {Hafezi}\ \emph {et~al.}(2011)\citenamefont {Hafezi},
  \citenamefont {Demler}, \citenamefont {Lukin},\ and\ \citenamefont
  {Taylor}}]{Hafezi11}%
  \BibitemOpen
  \bibfield  {author} {\bibinfo {author} {\bibfnamefont {M.}~\bibnamefont
  {Hafezi}}, \bibinfo {author} {\bibfnamefont {E.~A.}\ \bibnamefont {Demler}},
  \bibinfo {author} {\bibfnamefont {M.~D.}\ \bibnamefont {Lukin}}, \ and\
  \bibinfo {author} {\bibfnamefont {J.~M.}\ \bibnamefont {Taylor}},\ }\href
  {https://doi.org/10.1038/nphys2063} {\bibfield  {journal} {\bibinfo
  {journal} {Nat. Phys.}\ }\textbf {\bibinfo {volume} {7}},\ \bibinfo {pages}
  {907} (\bibinfo {year} {2011})}\BibitemShut {NoStop}%
\bibitem [{\citenamefont {Wang}\ \emph {et~al.}(2009)\citenamefont {Wang},
  \citenamefont {Chong}, \citenamefont {Joannopoulos},\ and\ \citenamefont
  {Soljacic}}]{Wang09}%
  \BibitemOpen
  \bibfield  {author} {\bibinfo {author} {\bibfnamefont {Z.}~\bibnamefont
  {Wang}}, \bibinfo {author} {\bibfnamefont {Y.}~\bibnamefont {Chong}},
  \bibinfo {author} {\bibfnamefont {J.~D.}\ \bibnamefont {Joannopoulos}}, \
  and\ \bibinfo {author} {\bibfnamefont {M.}~\bibnamefont {Soljacic}},\ }\href
  {https://doi.org/10.1038/nature08293} {\bibfield  {journal} {\bibinfo
  {journal} {Nature}\ }\textbf {\bibinfo {volume} {461}},\ \bibinfo {pages}
  {772 EP } (\bibinfo {year} {2009})}\BibitemShut {NoStop}%
\bibitem [{\citenamefont {Koch}\ \emph {et~al.}(2010)\citenamefont {Koch},
  \citenamefont {Houck}, \citenamefont {Hur},\ and\ \citenamefont
  {Girvin}}]{koch10}%
  \BibitemOpen
  \bibfield  {author} {\bibinfo {author} {\bibfnamefont {J.}~\bibnamefont
  {Koch}}, \bibinfo {author} {\bibfnamefont {A.~A.}\ \bibnamefont {Houck}},
  \bibinfo {author} {\bibfnamefont {K.~L.}\ \bibnamefont {Hur}}, \ and\
  \bibinfo {author} {\bibfnamefont {S.~M.}\ \bibnamefont {Girvin}},\ }\href
  {\doibase 10.1103/PhysRevA.82.043811} {\bibfield  {journal} {\bibinfo
  {journal} {Phys. Rev. A}\ }\textbf {\bibinfo {volume} {82}},\ \bibinfo
  {pages} {043811} (\bibinfo {year} {2010})}\BibitemShut {NoStop}%
\bibitem [{\citenamefont {Anderson}\ \emph {et~al.}(2016)\citenamefont
  {Anderson}, \citenamefont {Ma}, \citenamefont {Owens}, \citenamefont
  {Schuster},\ and\ \citenamefont {Simon}}]{Anderson16}%
  \BibitemOpen
  \bibfield  {author} {\bibinfo {author} {\bibfnamefont {B.~M.}\ \bibnamefont
  {Anderson}}, \bibinfo {author} {\bibfnamefont {R.}~\bibnamefont {Ma}},
  \bibinfo {author} {\bibfnamefont {C.}~\bibnamefont {Owens}}, \bibinfo
  {author} {\bibfnamefont {D.~I.}\ \bibnamefont {Schuster}}, \ and\ \bibinfo
  {author} {\bibfnamefont {J.}~\bibnamefont {Simon}},\ }\href {\doibase
  10.1103/PhysRevX.6.041043} {\bibfield  {journal} {\bibinfo  {journal} {Phys.
  Rev. X}\ }\textbf {\bibinfo {volume} {6}},\ \bibinfo {pages} {041043}
  (\bibinfo {year} {2016})}\BibitemShut {NoStop}%
\bibitem [{\citenamefont {Lu}\ \emph {et~al.}(2016)\citenamefont {Lu},
  \citenamefont {Fang}, \citenamefont {Fu}, \citenamefont {Johnson},
  \citenamefont {Joannopoulos},\ and\ \citenamefont {Soljacic}}]{Lu2016}%
  \BibitemOpen
  \bibfield  {author} {\bibinfo {author} {\bibfnamefont {L.}~\bibnamefont
  {Lu}}, \bibinfo {author} {\bibfnamefont {C.}~\bibnamefont {Fang}}, \bibinfo
  {author} {\bibfnamefont {L.}~\bibnamefont {Fu}}, \bibinfo {author}
  {\bibfnamefont {S.~G.}\ \bibnamefont {Johnson}}, \bibinfo {author}
  {\bibfnamefont {J.~D.}\ \bibnamefont {Joannopoulos}}, \ and\ \bibinfo
  {author} {\bibfnamefont {M.}~\bibnamefont {Soljacic}},\ }\href
  {https://doi.org/10.1038/nphys3611} {\bibfield  {journal} {\bibinfo
  {journal} {Nat. Phys.}\ }\textbf {\bibinfo {volume} {12}},\ \bibinfo {pages}
  {337} (\bibinfo {year} {2016})}\BibitemShut {NoStop}%
\bibitem [{\citenamefont {Owens}\ \emph {et~al.}(2018)\citenamefont {Owens},
  \citenamefont {LaChapelle}, \citenamefont {Saxberg}, \citenamefont
  {Anderson}, \citenamefont {Ma}, \citenamefont {Simon},\ and\ \citenamefont
  {Schuster}}]{Owens18}%
  \BibitemOpen
  \bibfield  {author} {\bibinfo {author} {\bibfnamefont {C.}~\bibnamefont
  {Owens}}, \bibinfo {author} {\bibfnamefont {A.}~\bibnamefont {LaChapelle}},
  \bibinfo {author} {\bibfnamefont {B.}~\bibnamefont {Saxberg}}, \bibinfo
  {author} {\bibfnamefont {B.~M.}\ \bibnamefont {Anderson}}, \bibinfo {author}
  {\bibfnamefont {R.}~\bibnamefont {Ma}}, \bibinfo {author} {\bibfnamefont
  {J.}~\bibnamefont {Simon}}, \ and\ \bibinfo {author} {\bibfnamefont {D.~I.}\
  \bibnamefont {Schuster}},\ }\href {\doibase 10.1103/PhysRevA.97.013818}
  {\bibfield  {journal} {\bibinfo  {journal} {Phys. Rev. A}\ }\textbf {\bibinfo
  {volume} {97}},\ \bibinfo {pages} {013818} (\bibinfo {year}
  {2018})}\BibitemShut {NoStop}%
\bibitem [{\citenamefont {Fang}\ \emph {et~al.}(2012)\citenamefont {Fang},
  \citenamefont {Yu},\ and\ \citenamefont {Fan}}]{fang12}%
  \BibitemOpen
  \bibfield  {author} {\bibinfo {author} {\bibfnamefont {K.}~\bibnamefont
  {Fang}}, \bibinfo {author} {\bibfnamefont {Z.}~\bibnamefont {Yu}}, \ and\
  \bibinfo {author} {\bibfnamefont {S.}~\bibnamefont {Fan}},\ }\href
  {https://www.nature.com/articles/nphoton.2012.236} {\bibfield  {journal}
  {\bibinfo  {journal} {Nat. Photon.}\ }\textbf {\bibinfo {volume} {6}},\
  \bibinfo {pages} {782} (\bibinfo {year} {2012})}\BibitemShut {NoStop}%
\bibitem [{\citenamefont {Peropadre}\ \emph {et~al.}(2013)\citenamefont
  {Peropadre}, \citenamefont {Zueco}, \citenamefont {Wulschner}, \citenamefont
  {Deppe}, \citenamefont {Marx}, \citenamefont {Gross},\ and\ \citenamefont
  {Garc\'{\i}a-Ripoll}}]{Peropadre13}%
  \BibitemOpen
  \bibfield  {author} {\bibinfo {author} {\bibfnamefont {B.}~\bibnamefont
  {Peropadre}}, \bibinfo {author} {\bibfnamefont {D.}~\bibnamefont {Zueco}},
  \bibinfo {author} {\bibfnamefont {F.}~\bibnamefont {Wulschner}}, \bibinfo
  {author} {\bibfnamefont {F.}~\bibnamefont {Deppe}}, \bibinfo {author}
  {\bibfnamefont {A.}~\bibnamefont {Marx}}, \bibinfo {author} {\bibfnamefont
  {R.}~\bibnamefont {Gross}}, \ and\ \bibinfo {author} {\bibfnamefont {J.~J.}\
  \bibnamefont {Garc\'{\i}a-Ripoll}},\ }\href {\doibase
  10.1103/PhysRevB.87.134504} {\bibfield  {journal} {\bibinfo  {journal} {Phys.
  Rev. B}\ }\textbf {\bibinfo {volume} {87}},\ \bibinfo {pages} {134504}
  (\bibinfo {year} {2013})}\BibitemShut {NoStop}%
\bibitem [{\citenamefont {Rechtsman}\ \emph {et~al.}(2013)\citenamefont
  {Rechtsman}, \citenamefont {Zeuner}, \citenamefont {Plotnik}, \citenamefont
  {Lumer}, \citenamefont {Podolsky}, \citenamefont {Dreisow}, \citenamefont
  {Nolte}, \citenamefont {Segev},\ and\ \citenamefont {Szameit}}]{Rechtsman13}%
  \BibitemOpen
  \bibfield  {author} {\bibinfo {author} {\bibfnamefont {M.~C.}\ \bibnamefont
  {Rechtsman}}, \bibinfo {author} {\bibfnamefont {J.~M.}\ \bibnamefont
  {Zeuner}}, \bibinfo {author} {\bibfnamefont {Y.}~\bibnamefont {Plotnik}},
  \bibinfo {author} {\bibfnamefont {Y.}~\bibnamefont {Lumer}}, \bibinfo
  {author} {\bibfnamefont {D.}~\bibnamefont {Podolsky}}, \bibinfo {author}
  {\bibfnamefont {F.}~\bibnamefont {Dreisow}}, \bibinfo {author} {\bibfnamefont
  {S.}~\bibnamefont {Nolte}}, \bibinfo {author} {\bibfnamefont
  {M.}~\bibnamefont {Segev}}, \ and\ \bibinfo {author} {\bibfnamefont
  {A.}~\bibnamefont {Szameit}},\ }\href {https://doi.org/10.1038/nature12066}
  {\bibfield  {journal} {\bibinfo  {journal} {Nature}\ }\textbf {\bibinfo
  {volume} {496}},\ \bibinfo {pages} {196} (\bibinfo {year}
  {2013})}\BibitemShut {NoStop}%
\bibitem [{\citenamefont {Roushan}\ \emph {et~al.}(2017)\citenamefont
  {Roushan}, \citenamefont {Neill}, \citenamefont {Megrant}, \citenamefont
  {Chen}, \citenamefont {Babbush}, \citenamefont {Barends}, \citenamefont
  {Campbell}, \citenamefont {Chen}, \citenamefont {Chiaro}, \citenamefont
  {Dunsworth} \emph {et~al.}}]{roushan17}%
  \BibitemOpen
  \bibfield  {author} {\bibinfo {author} {\bibfnamefont {P.}~\bibnamefont
  {Roushan}}, \bibinfo {author} {\bibfnamefont {C.}~\bibnamefont {Neill}},
  \bibinfo {author} {\bibfnamefont {A.}~\bibnamefont {Megrant}}, \bibinfo
  {author} {\bibfnamefont {Y.}~\bibnamefont {Chen}}, \bibinfo {author}
  {\bibfnamefont {R.}~\bibnamefont {Babbush}}, \bibinfo {author} {\bibfnamefont
  {R.}~\bibnamefont {Barends}}, \bibinfo {author} {\bibfnamefont
  {B.}~\bibnamefont {Campbell}}, \bibinfo {author} {\bibfnamefont
  {Z.}~\bibnamefont {Chen}}, \bibinfo {author} {\bibfnamefont {B.}~\bibnamefont
  {Chiaro}}, \bibinfo {author} {\bibfnamefont {A.}~\bibnamefont {Dunsworth}},
  \emph {et~al.},\ }\href {\doibase 10.1038/nphys3930} {\bibfield  {journal}
  {\bibinfo  {journal} {Nat. Phys.}\ }\textbf {\bibinfo {volume} {13}},\
  \bibinfo {pages} {146} (\bibinfo {year} {2017})}\BibitemShut {NoStop}%
\bibitem [{\citenamefont {Sounas}\ and\ \citenamefont
  {Al\`u}(2017)}]{sounas17}%
  \BibitemOpen
  \bibfield  {author} {\bibinfo {author} {\bibfnamefont {D.~L.}\ \bibnamefont
  {Sounas}}\ and\ \bibinfo {author} {\bibfnamefont {A.}~\bibnamefont {Al\`u}},\
  }\href {\doibase 10.1038/s41566-017-0051-x} {\bibfield  {journal} {\bibinfo
  {journal} {Nat. Photon.}\ }\textbf {\bibinfo {volume} {11}},\ \bibinfo
  {pages} {774} (\bibinfo {year} {2017})}\BibitemShut {NoStop}%
\bibitem [{\citenamefont {Mukherjee}\ \emph {et~al.}(2018)\citenamefont
  {Mukherjee}, \citenamefont {Di~Liberto}, \citenamefont {\"Ohberg},
  \citenamefont {Thomson},\ and\ \citenamefont {Goldman}}]{mukherjee18}%
  \BibitemOpen
  \bibfield  {author} {\bibinfo {author} {\bibfnamefont {S.}~\bibnamefont
  {Mukherjee}}, \bibinfo {author} {\bibfnamefont {M.}~\bibnamefont
  {Di~Liberto}}, \bibinfo {author} {\bibfnamefont {P.}~\bibnamefont
  {\"Ohberg}}, \bibinfo {author} {\bibfnamefont {R.~R.}\ \bibnamefont
  {Thomson}}, \ and\ \bibinfo {author} {\bibfnamefont {N.}~\bibnamefont
  {Goldman}},\ }\href {\doibase 10.1103/PhysRevLett.121.075502} {\bibfield
  {journal} {\bibinfo  {journal} {Phys. Rev. Lett.}\ }\textbf {\bibinfo
  {volume} {121}},\ \bibinfo {pages} {075502} (\bibinfo {year}
  {2018})}\BibitemShut {NoStop}%
\bibitem [{\citenamefont {Hafezi}\ and\ \citenamefont {Rabl}(2012)}]{Hafezi12}%
  \BibitemOpen
  \bibfield  {author} {\bibinfo {author} {\bibfnamefont {M.}~\bibnamefont
  {Hafezi}}\ and\ \bibinfo {author} {\bibfnamefont {P.}~\bibnamefont {Rabl}},\
  }\href {\doibase 10.1364/OE.20.007672} {\bibfield  {journal} {\bibinfo
  {journal} {Opt. Express}\ }\textbf {\bibinfo {volume} {20}},\ \bibinfo
  {pages} {7672} (\bibinfo {year} {2012})}\BibitemShut {NoStop}%
\bibitem [{\citenamefont {Schmidt}\ \emph {et~al.}(2015)\citenamefont
  {Schmidt}, \citenamefont {Kessler}, \citenamefont {Peano}, \citenamefont
  {Painter},\ and\ \citenamefont {Marquardt}}]{Schmidt15}%
  \BibitemOpen
  \bibfield  {author} {\bibinfo {author} {\bibfnamefont {M.}~\bibnamefont
  {Schmidt}}, \bibinfo {author} {\bibfnamefont {S.}~\bibnamefont {Kessler}},
  \bibinfo {author} {\bibfnamefont {V.}~\bibnamefont {Peano}}, \bibinfo
  {author} {\bibfnamefont {O.}~\bibnamefont {Painter}}, \ and\ \bibinfo
  {author} {\bibfnamefont {F.}~\bibnamefont {Marquardt}},\ }\href {\doibase
  10.1364/OPTICA.2.000635} {\bibfield  {journal} {\bibinfo  {journal} {Optica}\
  }\textbf {\bibinfo {volume} {2}},\ \bibinfo {pages} {635} (\bibinfo {year}
  {2015})}\BibitemShut {NoStop}%
\bibitem [{\citenamefont {{Ruesink}}\ \emph {et~al.}(2016)\citenamefont
  {{Ruesink}}, \citenamefont {{Miri}}, \citenamefont {{Al{\`u}}},\ and\
  \citenamefont {{Verhagen}}}]{Ruesink16}%
  \BibitemOpen
  \bibfield  {author} {\bibinfo {author} {\bibfnamefont {F.}~\bibnamefont
  {{Ruesink}}}, \bibinfo {author} {\bibfnamefont {M.-A.}\ \bibnamefont
  {{Miri}}}, \bibinfo {author} {\bibfnamefont {A.}~\bibnamefont {{Al{\`u}}}}, \
  and\ \bibinfo {author} {\bibfnamefont {E.}~\bibnamefont {{Verhagen}}},\
  }\href {\doibase 10.1038/ncomms13662} {\bibfield  {journal} {\bibinfo
  {journal} {Nat. Commun.}\ }\textbf {\bibinfo {volume} {7}},\ \bibinfo {eid}
  {13662} (\bibinfo {year} {2016})}\BibitemShut {NoStop}%
\bibitem [{\citenamefont {{Shen}}\ \emph {et~al.}(2016)\citenamefont {{Shen}},
  \citenamefont {{Zhang}}, \citenamefont {{Chen}}, \citenamefont {{Zou}},
  \citenamefont {{Xiao}}, \citenamefont {{Zou}}, \citenamefont {{Sun}},
  \citenamefont {{Guo}},\ and\ \citenamefont {{Dong}}}]{Shen16}%
  \BibitemOpen
  \bibfield  {author} {\bibinfo {author} {\bibfnamefont {Z.}~\bibnamefont
  {{Shen}}}, \bibinfo {author} {\bibfnamefont {Y.-L.}\ \bibnamefont {{Zhang}}},
  \bibinfo {author} {\bibfnamefont {Y.}~\bibnamefont {{Chen}}}, \bibinfo
  {author} {\bibfnamefont {C.-L.}\ \bibnamefont {{Zou}}}, \bibinfo {author}
  {\bibfnamefont {Y.-F.}\ \bibnamefont {{Xiao}}}, \bibinfo {author}
  {\bibfnamefont {X.-B.}\ \bibnamefont {{Zou}}}, \bibinfo {author}
  {\bibfnamefont {F.-W.}\ \bibnamefont {{Sun}}}, \bibinfo {author}
  {\bibfnamefont {G.-C.}\ \bibnamefont {{Guo}}}, \ and\ \bibinfo {author}
  {\bibfnamefont {C.-H.}\ \bibnamefont {{Dong}}},\ }\href {\doibase
  10.1038/nphoton.2016.161} {\bibfield  {journal} {\bibinfo  {journal} {Nat.
  Photon.}\ }\textbf {\bibinfo {volume} {10}},\ \bibinfo {pages} {657}
  (\bibinfo {year} {2016})}\BibitemShut {NoStop}%
\bibitem [{\citenamefont {Bernier}\ \emph {et~al.}(2018)\citenamefont
  {Bernier}, \citenamefont {T\'oth}, \citenamefont {Feofanov},\ and\
  \citenamefont {Kippenberg}}]{bernier18pra}%
  \BibitemOpen
  \bibfield  {author} {\bibinfo {author} {\bibfnamefont {N.~R.}\ \bibnamefont
  {Bernier}}, \bibinfo {author} {\bibfnamefont {L.~D.}\ \bibnamefont {T\'oth}},
  \bibinfo {author} {\bibfnamefont {A.~K.}\ \bibnamefont {Feofanov}}, \ and\
  \bibinfo {author} {\bibfnamefont {T.~J.}\ \bibnamefont {Kippenberg}},\ }\href
  {\doibase 10.1103/PhysRevA.98.023841} {\bibfield  {journal} {\bibinfo
  {journal} {Phys. Rev. A}\ }\textbf {\bibinfo {volume} {98}},\ \bibinfo
  {pages} {023841} (\bibinfo {year} {2018})}\BibitemShut {NoStop}%
\bibitem [{\citenamefont {Bermudez}\ \emph {et~al.}(2011)\citenamefont
  {Bermudez}, \citenamefont {Schaetz},\ and\ \citenamefont
  {Porras}}]{Bermudez11prl}%
  \BibitemOpen
  \bibfield  {author} {\bibinfo {author} {\bibfnamefont {A.}~\bibnamefont
  {Bermudez}}, \bibinfo {author} {\bibfnamefont {T.}~\bibnamefont {Schaetz}}, \
  and\ \bibinfo {author} {\bibfnamefont {D.}~\bibnamefont {Porras}},\ }\href
  {\doibase 10.1103/PhysRevLett.107.150501} {\bibfield  {journal} {\bibinfo
  {journal} {Phys. Rev. Lett.}\ }\textbf {\bibinfo {volume} {107}},\ \bibinfo
  {pages} {150501} (\bibinfo {year} {2011})}\BibitemShut {NoStop}%
\bibitem [{\citenamefont {Bermudez}\ \emph {et~al.}(2012)\citenamefont
  {Bermudez}, \citenamefont {Schaetz},\ and\ \citenamefont
  {Porras}}]{Bermudez12}%
  \BibitemOpen
  \bibfield  {author} {\bibinfo {author} {\bibfnamefont {A.}~\bibnamefont
  {Bermudez}}, \bibinfo {author} {\bibfnamefont {T.}~\bibnamefont {Schaetz}}, \
  and\ \bibinfo {author} {\bibfnamefont {D.}~\bibnamefont {Porras}},\ }\href
  {http://stacks.iop.org/1367-2630/14/i=5/a=053049} {\bibfield  {journal}
  {\bibinfo  {journal} {New J. Phys.}\ }\textbf {\bibinfo {volume} {14}},\
  \bibinfo {pages} {053049} (\bibinfo {year} {2012})}\BibitemShut {NoStop}%
\bibitem [{\citenamefont {S{\"u}sstrunk}\ and\ \citenamefont
  {Huber}(2015)}]{Susstrunk15}%
  \BibitemOpen
  \bibfield  {author} {\bibinfo {author} {\bibfnamefont {R.}~\bibnamefont
  {S{\"u}sstrunk}}\ and\ \bibinfo {author} {\bibfnamefont {S.~D.}\ \bibnamefont
  {Huber}},\ }\href {\doibase 10.1126/science.aab0239} {\bibfield  {journal}
  {\bibinfo  {journal} {Science}\ }\textbf {\bibinfo {volume} {349}},\ \bibinfo
  {pages} {47} (\bibinfo {year} {2015})}\BibitemShut {NoStop}%
\bibitem [{\citenamefont {Harder}\ \emph {et~al.}(2017)\citenamefont {Harder},
  \citenamefont {Bai}, \citenamefont {Hyde},\ and\ \citenamefont
  {Hu}}]{harder17prb}%
  \BibitemOpen
  \bibfield  {author} {\bibinfo {author} {\bibfnamefont {M.}~\bibnamefont
  {Harder}}, \bibinfo {author} {\bibfnamefont {L.}~\bibnamefont {Bai}},
  \bibinfo {author} {\bibfnamefont {P.}~\bibnamefont {Hyde}}, \ and\ \bibinfo
  {author} {\bibfnamefont {C.-M.}\ \bibnamefont {Hu}},\ }\href {\doibase
  10.1103/PhysRevB.95.214411} {\bibfield  {journal} {\bibinfo  {journal} {Phys.
  Rev. B}\ }\textbf {\bibinfo {volume} {95}},\ \bibinfo {pages} {214411}
  (\bibinfo {year} {2017})}\BibitemShut {NoStop}%
\bibitem [{\citenamefont {Zhang}\ \emph {et~al.}(2017)\citenamefont {Zhang},
  \citenamefont {Luo}, \citenamefont {Wang}, \citenamefont {Li},\ and\
  \citenamefont {You}}]{zhang17natcomm}%
  \BibitemOpen
  \bibfield  {author} {\bibinfo {author} {\bibfnamefont {D.}~\bibnamefont
  {Zhang}}, \bibinfo {author} {\bibfnamefont {X.-Q.}\ \bibnamefont {Luo}},
  \bibinfo {author} {\bibfnamefont {Y.-P.}\ \bibnamefont {Wang}}, \bibinfo
  {author} {\bibfnamefont {T.-F.}\ \bibnamefont {Li}}, \ and\ \bibinfo {author}
  {\bibfnamefont {J.~Q.}\ \bibnamefont {You}},\ }\href {\doibase
  10.1038/s41467-017-01634-w} {\bibfield  {journal} {\bibinfo  {journal}
  {Nature Communications}\ }\textbf {\bibinfo {volume} {8}},\ \bibinfo {pages}
  {1368} (\bibinfo {year} {2017})}\BibitemShut {NoStop}%
\bibitem [{\citenamefont {{Houck}}\ \emph {et~al.}(2012)\citenamefont
  {{Houck}}, \citenamefont {{T{\"u}reci}},\ and\ \citenamefont
  {{Koch}}}]{houck12natphys}%
  \BibitemOpen
  \bibfield  {author} {\bibinfo {author} {\bibfnamefont {A.~A.}\ \bibnamefont
  {{Houck}}}, \bibinfo {author} {\bibfnamefont {H.~E.}\ \bibnamefont
  {{T{\"u}reci}}}, \ and\ \bibinfo {author} {\bibfnamefont {J.}~\bibnamefont
  {{Koch}}},\ }\href
  {http://www.nature.com/nphys/journal/v8/n4/abs/nphys2251.html} {\bibfield
  {journal} {\bibinfo  {journal} {Nature Phys.}\ }\textbf {\bibinfo {volume}
  {8}} (\bibinfo {year} {2012})}\BibitemShut {NoStop}%
\bibitem [{\citenamefont {Hartmann}(2016)}]{Hartmann16}%
  \BibitemOpen
  \bibfield  {author} {\bibinfo {author} {\bibfnamefont {M.~J.}\ \bibnamefont
  {Hartmann}},\ }\href {http://stacks.iop.org/2040-8986/18/i=10/a=104005}
  {\bibfield  {journal} {\bibinfo  {journal} {J. Opt.}\ }\textbf {\bibinfo
  {volume} {18}},\ \bibinfo {pages} {104005} (\bibinfo {year}
  {2016})}\BibitemShut {NoStop}%
\bibitem [{\citenamefont {Longhi}(2017)}]{Longhi17}%
  \BibitemOpen
  \bibfield  {author} {\bibinfo {author} {\bibfnamefont {S.}~\bibnamefont
  {Longhi}},\ }\href {http://stacks.iop.org/0295-5075/120/i=6/a=64001}
  {\bibfield  {journal} {\bibinfo  {journal} {EPL}\ }\textbf {\bibinfo {volume}
  {120}},\ \bibinfo {pages} {64001} (\bibinfo {year} {2017})}\BibitemShut
  {NoStop}%
\bibitem [{\citenamefont {Peano}\ \emph {et~al.}(2016)\citenamefont {Peano},
  \citenamefont {Houde}, \citenamefont {Marquardt},\ and\ \citenamefont
  {Clerk}}]{peano16prx}%
  \BibitemOpen
  \bibfield  {author} {\bibinfo {author} {\bibfnamefont {V.}~\bibnamefont
  {Peano}}, \bibinfo {author} {\bibfnamefont {M.}~\bibnamefont {Houde}},
  \bibinfo {author} {\bibfnamefont {F.}~\bibnamefont {Marquardt}}, \ and\
  \bibinfo {author} {\bibfnamefont {A.~A.}\ \bibnamefont {Clerk}},\ }\href
  {\doibase 10.1103/PhysRevX.6.041026} {\bibfield  {journal} {\bibinfo
  {journal} {Phys. Rev. X}\ }\textbf {\bibinfo {volume} {6}},\ \bibinfo {pages}
  {041026} (\bibinfo {year} {2016})}\BibitemShut {NoStop}%
\bibitem [{\citenamefont {Esaki}\ \emph {et~al.}(2011)\citenamefont {Esaki},
  \citenamefont {Sato}, \citenamefont {Hasebe},\ and\ \citenamefont
  {Kohmoto}}]{Esaki11}%
  \BibitemOpen
  \bibfield  {author} {\bibinfo {author} {\bibfnamefont {K.}~\bibnamefont
  {Esaki}}, \bibinfo {author} {\bibfnamefont {M.}~\bibnamefont {Sato}},
  \bibinfo {author} {\bibfnamefont {K.}~\bibnamefont {Hasebe}}, \ and\ \bibinfo
  {author} {\bibfnamefont {M.}~\bibnamefont {Kohmoto}},\ }\href {\doibase
  10.1103/PhysRevB.84.205128} {\bibfield  {journal} {\bibinfo  {journal} {Phys.
  Rev. B}\ }\textbf {\bibinfo {volume} {84}},\ \bibinfo {pages} {205128}
  (\bibinfo {year} {2011})}\BibitemShut {NoStop}%
\bibitem [{\citenamefont {Schomerus}(2013)}]{Schomerus13}%
  \BibitemOpen
  \bibfield  {author} {\bibinfo {author} {\bibfnamefont {H.}~\bibnamefont
  {Schomerus}},\ }\href {http://ol.osa.org/abstract.cfm?URI=ol-38-11-1912}
  {\bibfield  {journal} {\bibinfo  {journal} {Opt. Lett.}\ }\textbf {\bibinfo
  {volume} {38}},\ \bibinfo {pages} {1912} (\bibinfo {year}
  {2013})}\BibitemShut {NoStop}%
\bibitem [{\citenamefont {Lee}(2016)}]{Lee16}%
  \BibitemOpen
  \bibfield  {author} {\bibinfo {author} {\bibfnamefont {T.~E.}\ \bibnamefont
  {Lee}},\ }\href {\doibase 10.1103/PhysRevLett.116.133903} {\bibfield
  {journal} {\bibinfo  {journal} {Phys. Rev. Lett.}\ }\textbf {\bibinfo
  {volume} {116}},\ \bibinfo {pages} {133903} (\bibinfo {year}
  {2016})}\BibitemShut {NoStop}%
\bibitem [{\citenamefont {Leykam}\ \emph {et~al.}(2017)\citenamefont {Leykam},
  \citenamefont {Bliokh}, \citenamefont {Huang}, \citenamefont {Chong},\ and\
  \citenamefont {Nori}}]{Leykam17}%
  \BibitemOpen
  \bibfield  {author} {\bibinfo {author} {\bibfnamefont {D.}~\bibnamefont
  {Leykam}}, \bibinfo {author} {\bibfnamefont {K.~Y.}\ \bibnamefont {Bliokh}},
  \bibinfo {author} {\bibfnamefont {C.}~\bibnamefont {Huang}}, \bibinfo
  {author} {\bibfnamefont {Y.~D.}\ \bibnamefont {Chong}}, \ and\ \bibinfo
  {author} {\bibfnamefont {F.}~\bibnamefont {Nori}},\ }\href {\doibase
  10.1103/PhysRevLett.118.040401} {\bibfield  {journal} {\bibinfo  {journal}
  {Phys. Rev. Lett.}\ }\textbf {\bibinfo {volume} {118}},\ \bibinfo {pages}
  {040401} (\bibinfo {year} {2017})}\BibitemShut {NoStop}%
\bibitem [{\citenamefont {Gong}\ \emph {et~al.}(2018)\citenamefont {Gong},
  \citenamefont {Ashida}, \citenamefont {Kawabata}, \citenamefont {Takasan},
  \citenamefont {Higashikawa},\ and\ \citenamefont {Ueda}}]{Gong18}%
  \BibitemOpen
  \bibfield  {author} {\bibinfo {author} {\bibfnamefont {Z.}~\bibnamefont
  {Gong}}, \bibinfo {author} {\bibfnamefont {Y.}~\bibnamefont {Ashida}},
  \bibinfo {author} {\bibfnamefont {K.}~\bibnamefont {Kawabata}}, \bibinfo
  {author} {\bibfnamefont {K.}~\bibnamefont {Takasan}}, \bibinfo {author}
  {\bibfnamefont {S.}~\bibnamefont {Higashikawa}}, \ and\ \bibinfo {author}
  {\bibfnamefont {M.}~\bibnamefont {Ueda}},\ }\href {\doibase
  10.1103/PhysRevX.8.031079} {\bibfield  {journal} {\bibinfo  {journal} {Phys.
  Rev. X}\ }\textbf {\bibinfo {volume} {8}},\ \bibinfo {pages} {031079}
  (\bibinfo {year} {2018})}\BibitemShut {NoStop}%
\bibitem [{\citenamefont {Shen}\ \emph {et~al.}(2018)\citenamefont {Shen},
  \citenamefont {Zhen},\ and\ \citenamefont {Fu}}]{shen18}%
  \BibitemOpen
  \bibfield  {author} {\bibinfo {author} {\bibfnamefont {H.}~\bibnamefont
  {Shen}}, \bibinfo {author} {\bibfnamefont {B.}~\bibnamefont {Zhen}}, \ and\
  \bibinfo {author} {\bibfnamefont {L.}~\bibnamefont {Fu}},\ }\href {\doibase
  10.1103/PhysRevLett.120.146402} {\bibfield  {journal} {\bibinfo  {journal}
  {Phys. Rev. Lett.}\ }\textbf {\bibinfo {volume} {120}},\ \bibinfo {pages}
  {146402} (\bibinfo {year} {2018})}\BibitemShut {NoStop}%
\bibitem [{\citenamefont {Poli}\ \emph {et~al.}(2015)\citenamefont {Poli},
  \citenamefont {Bellec}, \citenamefont {Kuhl}, \citenamefont {Mortessagne},\
  and\ \citenamefont {Schomerus}}]{Poli15}%
  \BibitemOpen
  \bibfield  {author} {\bibinfo {author} {\bibfnamefont {C.}~\bibnamefont
  {Poli}}, \bibinfo {author} {\bibfnamefont {M.}~\bibnamefont {Bellec}},
  \bibinfo {author} {\bibfnamefont {U.}~\bibnamefont {Kuhl}}, \bibinfo {author}
  {\bibfnamefont {F.}~\bibnamefont {Mortessagne}}, \ and\ \bibinfo {author}
  {\bibfnamefont {H.}~\bibnamefont {Schomerus}},\ }\href
  {https://doi.org/10.1038/ncomms7710} {\bibfield  {journal} {\bibinfo
  {journal} {Nat. Commun.}\ }\textbf {\bibinfo {volume} {6}},\ \bibinfo {pages}
  {6710} (\bibinfo {year} {2015})}\BibitemShut {NoStop}%
\bibitem [{\citenamefont {Zeuner}\ \emph {et~al.}(2015)\citenamefont {Zeuner},
  \citenamefont {Rechtsman}, \citenamefont {Plotnik}, \citenamefont {Lumer},
  \citenamefont {Nolte}, \citenamefont {Rudner}, \citenamefont {Segev},\ and\
  \citenamefont {Szameit}}]{Zeuner15}%
  \BibitemOpen
  \bibfield  {author} {\bibinfo {author} {\bibfnamefont {J.~M.}\ \bibnamefont
  {Zeuner}}, \bibinfo {author} {\bibfnamefont {M.~C.}\ \bibnamefont
  {Rechtsman}}, \bibinfo {author} {\bibfnamefont {Y.}~\bibnamefont {Plotnik}},
  \bibinfo {author} {\bibfnamefont {Y.}~\bibnamefont {Lumer}}, \bibinfo
  {author} {\bibfnamefont {S.}~\bibnamefont {Nolte}}, \bibinfo {author}
  {\bibfnamefont {M.~S.}\ \bibnamefont {Rudner}}, \bibinfo {author}
  {\bibfnamefont {M.}~\bibnamefont {Segev}}, \ and\ \bibinfo {author}
  {\bibfnamefont {A.}~\bibnamefont {Szameit}},\ }\href {\doibase
  10.1103/PhysRevLett.115.040402} {\bibfield  {journal} {\bibinfo  {journal}
  {Phys. Rev. Lett.}\ }\textbf {\bibinfo {volume} {115}},\ \bibinfo {pages}
  {040402} (\bibinfo {year} {2015})}\BibitemShut {NoStop}%
\bibitem [{\citenamefont {Pan}\ \emph {et~al.}(2018)\citenamefont {Pan},
  \citenamefont {Zhao}, \citenamefont {Miao}, \citenamefont {Longhi},\ and\
  \citenamefont {Feng}}]{Pan18}%
  \BibitemOpen
  \bibfield  {author} {\bibinfo {author} {\bibfnamefont {M.}~\bibnamefont
  {Pan}}, \bibinfo {author} {\bibfnamefont {P.}~\bibnamefont {Zhao}}, \bibinfo
  {author} {\bibfnamefont {P.}~\bibnamefont {Miao}}, \bibinfo {author}
  {\bibfnamefont {S.}~\bibnamefont {Longhi}}, \ and\ \bibinfo {author}
  {\bibfnamefont {L.}~\bibnamefont {Feng}},\ }\href
  {https://www.nature.com/articles/s41467-018-03822-8} {\bibfield  {journal}
  {\bibinfo  {journal} {Nat. Commun.}\ }\textbf {\bibinfo {volume} {9}},\
  \bibinfo {pages} {1308} (\bibinfo {year} {2018})}\BibitemShut {NoStop}%
\bibitem [{\citenamefont {Hadad}\ \emph {et~al.}(2018)\citenamefont {Hadad},
  \citenamefont {Soric}, \citenamefont {Khanikaev},\ and\ \citenamefont
  {Al{\`u}}}]{hadad18}%
  \BibitemOpen
  \bibfield  {author} {\bibinfo {author} {\bibfnamefont {Y.}~\bibnamefont
  {Hadad}}, \bibinfo {author} {\bibfnamefont {J.~C.}\ \bibnamefont {Soric}},
  \bibinfo {author} {\bibfnamefont {A.~B.}\ \bibnamefont {Khanikaev}}, \ and\
  \bibinfo {author} {\bibfnamefont {A.}~\bibnamefont {Al{\`u}}},\ }\href
  {https://www.nature.com/articles/s41928-018-0042-z} {\bibfield  {journal}
  {\bibinfo  {journal} {Nat. Electron.}\ }\textbf {\bibinfo {volume} {1}},\
  \bibinfo {pages} {178} (\bibinfo {year} {2018})}\BibitemShut {NoStop}%
\bibitem [{\citenamefont {Fang}\ \emph {et~al.}(2017)\citenamefont {Fang},
  \citenamefont {Luo}, \citenamefont {Metelmann}, \citenamefont {Matheny},
  \citenamefont {Marquardt}, \citenamefont {Clerk},\ and\ \citenamefont
  {Painter}}]{fang17}%
  \BibitemOpen
  \bibfield  {author} {\bibinfo {author} {\bibfnamefont {K.}~\bibnamefont
  {Fang}}, \bibinfo {author} {\bibfnamefont {J.}~\bibnamefont {Luo}}, \bibinfo
  {author} {\bibfnamefont {A.}~\bibnamefont {Metelmann}}, \bibinfo {author}
  {\bibfnamefont {M.~H.}\ \bibnamefont {Matheny}}, \bibinfo {author}
  {\bibfnamefont {F.}~\bibnamefont {Marquardt}}, \bibinfo {author}
  {\bibfnamefont {A.~A.}\ \bibnamefont {Clerk}}, \ and\ \bibinfo {author}
  {\bibfnamefont {O.}~\bibnamefont {Painter}},\ }\href {\doibase
  10.1038/nphys4009} {\bibfield  {journal} {\bibinfo  {journal} {Nat. Phys.}\
  }\textbf {\bibinfo {volume} {13}},\ \bibinfo {pages} {465} (\bibinfo {year}
  {2017})}\BibitemShut {NoStop}%
\bibitem [{\citenamefont {Ryu}\ and\ \citenamefont {Hatsugai}(2002)}]{ryu02}%
  \BibitemOpen
  \bibfield  {author} {\bibinfo {author} {\bibfnamefont {S.}~\bibnamefont
  {Ryu}}\ and\ \bibinfo {author} {\bibfnamefont {Y.}~\bibnamefont {Hatsugai}},\
  }\href {\doibase 10.1103/PhysRevLett.89.077002} {\bibfield  {journal}
  {\bibinfo  {journal} {Phys. Rev. Lett.}\ }\textbf {\bibinfo {volume} {89}},\
  \bibinfo {pages} {077002} (\bibinfo {year} {2002})}\BibitemShut {NoStop}%
\bibitem [{\citenamefont {Hatano}\ and\ \citenamefont
  {Nelson}(1996)}]{Hatano96}%
  \BibitemOpen
  \bibfield  {author} {\bibinfo {author} {\bibfnamefont {N.}~\bibnamefont
  {Hatano}}\ and\ \bibinfo {author} {\bibfnamefont {D.~R.}\ \bibnamefont
  {Nelson}},\ }\href {\doibase 10.1103/PhysRevLett.77.570} {\bibfield
  {journal} {\bibinfo  {journal} {Phys. Rev. Lett.}\ }\textbf {\bibinfo
  {volume} {77}},\ \bibinfo {pages} {570} (\bibinfo {year} {1996})}\BibitemShut
  {NoStop}%
\bibitem [{\citenamefont {Longhi}\ \emph {et~al.}(2015)\citenamefont {Longhi},
  \citenamefont {Gatti},\ and\ \citenamefont {Della~Valle}}]{Longhi15}%
  \BibitemOpen
  \bibfield  {author} {\bibinfo {author} {\bibfnamefont {S.}~\bibnamefont
  {Longhi}}, \bibinfo {author} {\bibfnamefont {D.}~\bibnamefont {Gatti}}, \
  and\ \bibinfo {author} {\bibfnamefont {G.}~\bibnamefont {Della~Valle}},\
  }\href {\doibase 10.1103/PhysRevB.92.094204} {\bibfield  {journal} {\bibinfo
  {journal} {Phys. Rev. B}\ }\textbf {\bibinfo {volume} {92}},\ \bibinfo
  {pages} {094204} (\bibinfo {year} {2015})}\BibitemShut {NoStop}%
\bibitem [{\citenamefont {Asb{\'o}th}\ \emph {et~al.}(2016)\citenamefont
  {Asb{\'o}th}, \citenamefont {Oroszl{\'a}ny},\ and\ \citenamefont
  {P{\'a}lyi}}]{asboth16}%
  \BibitemOpen
  \bibfield  {author} {\bibinfo {author} {\bibfnamefont {J.~K.}\ \bibnamefont
  {Asb{\'o}th}}, \bibinfo {author} {\bibfnamefont {L.}~\bibnamefont
  {Oroszl{\'a}ny}}, \ and\ \bibinfo {author} {\bibfnamefont {A.}~\bibnamefont
  {P{\'a}lyi}},\ }\href@noop {} {\bibfield  {journal} {\bibinfo  {journal}
  {Lecture Notes in Physics}\ }\textbf {\bibinfo {volume} {919}} (\bibinfo
  {year} {2016})}\BibitemShut {NoStop}%
\bibitem [{\citenamefont {Heeger}\ \emph {et~al.}(1988)\citenamefont {Heeger},
  \citenamefont {Kivelson}, \citenamefont {Schrieffer},\ and\ \citenamefont
  {Su}}]{Heeger88}%
  \BibitemOpen
  \bibfield  {author} {\bibinfo {author} {\bibfnamefont {A.~J.}\ \bibnamefont
  {Heeger}}, \bibinfo {author} {\bibfnamefont {S.}~\bibnamefont {Kivelson}},
  \bibinfo {author} {\bibfnamefont {J.~R.}\ \bibnamefont {Schrieffer}}, \ and\
  \bibinfo {author} {\bibfnamefont {W.~P.}\ \bibnamefont {Su}},\ }\href
  {\doibase 10.1103/RevModPhys.60.781} {\bibfield  {journal} {\bibinfo
  {journal} {Rev. Mod. Phys.}\ }\textbf {\bibinfo {volume} {60}},\ \bibinfo
  {pages} {781} (\bibinfo {year} {1988})}\BibitemShut {NoStop}%
\bibitem [{\citenamefont {Campos~Venuti}\ \emph {et~al.}(2007)\citenamefont
  {Campos~Venuti}, \citenamefont {Giampaolo}, \citenamefont {Illuminati},\ and\
  \citenamefont {Zanardi}}]{Venuti07}%
  \BibitemOpen
  \bibfield  {author} {\bibinfo {author} {\bibfnamefont {L.}~\bibnamefont
  {Campos~Venuti}}, \bibinfo {author} {\bibfnamefont {S.~M.}\ \bibnamefont
  {Giampaolo}}, \bibinfo {author} {\bibfnamefont {F.}~\bibnamefont
  {Illuminati}}, \ and\ \bibinfo {author} {\bibfnamefont {P.}~\bibnamefont
  {Zanardi}},\ }\href {\doibase 10.1103/PhysRevA.76.052328} {\bibfield
  {journal} {\bibinfo  {journal} {Phys. Rev. A}\ }\textbf {\bibinfo {volume}
  {76}},\ \bibinfo {pages} {052328} (\bibinfo {year} {2007})}\BibitemShut
  {NoStop}%
\bibitem [{\citenamefont {Noschese}\ \emph {et~al.}(2013)\citenamefont
  {Noschese}, \citenamefont {Pasquini},\ and\ \citenamefont
  {Reichel}}]{noschese13}%
  \BibitemOpen
  \bibfield  {author} {\bibinfo {author} {\bibfnamefont {S.}~\bibnamefont
  {Noschese}}, \bibinfo {author} {\bibfnamefont {L.}~\bibnamefont {Pasquini}},
  \ and\ \bibinfo {author} {\bibfnamefont {L.}~\bibnamefont {Reichel}},\
  }\href@noop {} {\bibfield  {journal} {\bibinfo  {journal} {Numerical linear
  algebra with applications}\ }\textbf {\bibinfo {volume} {20}},\ \bibinfo
  {pages} {302} (\bibinfo {year} {2013})}\BibitemShut {NoStop}%
\bibitem [{\citenamefont {{Kamal}}\ \emph {et~al.}(2011)\citenamefont
  {{Kamal}}, \citenamefont {{Clarke}},\ and\ \citenamefont
  {{Devoret}}}]{Devoret11}%
  \BibitemOpen
  \bibfield  {author} {\bibinfo {author} {\bibfnamefont {A.}~\bibnamefont
  {{Kamal}}}, \bibinfo {author} {\bibfnamefont {J.}~\bibnamefont {{Clarke}}}, \
  and\ \bibinfo {author} {\bibfnamefont {M.~H.}\ \bibnamefont {{Devoret}}},\
  }\href {\doibase 10.1038/nphys1893} {\bibfield  {journal} {\bibinfo
  {journal} {Nat. Phys.}\ }\textbf {\bibinfo {volume} {7}},\ \bibinfo {pages}
  {311} (\bibinfo {year} {2011})}\BibitemShut {NoStop}%
\bibitem [{\citenamefont {Abdo}\ \emph {et~al.}(2013)\citenamefont {Abdo},
  \citenamefont {Sliwa}, \citenamefont {Frunzio},\ and\ \citenamefont
  {Devoret}}]{Devoret13}%
  \BibitemOpen
  \bibfield  {author} {\bibinfo {author} {\bibfnamefont {B.}~\bibnamefont
  {Abdo}}, \bibinfo {author} {\bibfnamefont {K.}~\bibnamefont {Sliwa}},
  \bibinfo {author} {\bibfnamefont {L.}~\bibnamefont {Frunzio}}, \ and\
  \bibinfo {author} {\bibfnamefont {M.}~\bibnamefont {Devoret}},\ }\href
  {\doibase 10.1103/PhysRevX.3.031001} {\bibfield  {journal} {\bibinfo
  {journal} {Phys. Rev. X}\ }\textbf {\bibinfo {volume} {3}},\ \bibinfo {pages}
  {031001} (\bibinfo {year} {2013})}\BibitemShut {NoStop}%
\bibitem [{\citenamefont {Sliwa}\ \emph {et~al.}(2015)\citenamefont {Sliwa},
  \citenamefont {Hatridge}, \citenamefont {Narla}, \citenamefont {Shankar},
  \citenamefont {Frunzio}, \citenamefont {Schoelkopf},\ and\ \citenamefont
  {Devoret}}]{Devoret15}%
  \BibitemOpen
  \bibfield  {author} {\bibinfo {author} {\bibfnamefont {K.~M.}\ \bibnamefont
  {Sliwa}}, \bibinfo {author} {\bibfnamefont {M.}~\bibnamefont {Hatridge}},
  \bibinfo {author} {\bibfnamefont {A.}~\bibnamefont {Narla}}, \bibinfo
  {author} {\bibfnamefont {S.}~\bibnamefont {Shankar}}, \bibinfo {author}
  {\bibfnamefont {L.}~\bibnamefont {Frunzio}}, \bibinfo {author} {\bibfnamefont
  {R.~J.}\ \bibnamefont {Schoelkopf}}, \ and\ \bibinfo {author} {\bibfnamefont
  {M.~H.}\ \bibnamefont {Devoret}},\ }\href {\doibase
  10.1103/PhysRevX.5.041020} {\bibfield  {journal} {\bibinfo  {journal} {Phys.
  Rev. X}\ }\textbf {\bibinfo {volume} {5}},\ \bibinfo {pages} {041020}
  (\bibinfo {year} {2015})}\BibitemShut {NoStop}%
\bibitem [{\citenamefont {Metelmann}\ and\ \citenamefont
  {Clerk}(2015)}]{Metelmann15}%
  \BibitemOpen
  \bibfield  {author} {\bibinfo {author} {\bibfnamefont {A.}~\bibnamefont
  {Metelmann}}\ and\ \bibinfo {author} {\bibfnamefont {A.~A.}\ \bibnamefont
  {Clerk}},\ }\href {\doibase 10.1103/PhysRevX.5.021025} {\bibfield  {journal}
  {\bibinfo  {journal} {Phys. Rev. X}\ }\textbf {\bibinfo {volume} {5}},\
  \bibinfo {pages} {021025} (\bibinfo {year} {2015})}\BibitemShut {NoStop}%
\bibitem [{\citenamefont {Malz}\ \emph {et~al.}(2018)\citenamefont {Malz},
  \citenamefont {T\'oth}, \citenamefont {Bernier}, \citenamefont {Feofanov},
  \citenamefont {Kippenberg},\ and\ \citenamefont {Nunnenkamp}}]{Nunnenkamp18}%
  \BibitemOpen
  \bibfield  {author} {\bibinfo {author} {\bibfnamefont {D.}~\bibnamefont
  {Malz}}, \bibinfo {author} {\bibfnamefont {L.~D.}\ \bibnamefont {T\'oth}},
  \bibinfo {author} {\bibfnamefont {N.~R.}\ \bibnamefont {Bernier}}, \bibinfo
  {author} {\bibfnamefont {A.~K.}\ \bibnamefont {Feofanov}}, \bibinfo {author}
  {\bibfnamefont {T.~J.}\ \bibnamefont {Kippenberg}}, \ and\ \bibinfo {author}
  {\bibfnamefont {A.}~\bibnamefont {Nunnenkamp}},\ }\href {\doibase
  10.1103/PhysRevLett.120.023601} {\bibfield  {journal} {\bibinfo  {journal}
  {Phys. Rev. Lett.}\ }\textbf {\bibinfo {volume} {120}},\ \bibinfo {pages}
  {023601} (\bibinfo {year} {2018})}\BibitemShut {NoStop}%
\bibitem [{\citenamefont {Quijandria}\ \emph {et~al.}(2013)\citenamefont
  {Quijandria}, \citenamefont {Porras}, \citenamefont {Garcia-Ripoll},\ and\
  \citenamefont {Zueco}}]{Quijandria13prl}%
  \BibitemOpen
  \bibfield  {author} {\bibinfo {author} {\bibfnamefont {F.}~\bibnamefont
  {Quijandria}}, \bibinfo {author} {\bibfnamefont {D.}~\bibnamefont {Porras}},
  \bibinfo {author} {\bibfnamefont {J.~J.}\ \bibnamefont {Garcia-Ripoll}}, \
  and\ \bibinfo {author} {\bibfnamefont {D.}~\bibnamefont {Zueco}},\ }\href
  {\doibase 10.1103/PhysRevLett.111.073602} {\bibfield  {journal} {\bibinfo
  {journal} {Phys. Rev. Lett.}\ }\textbf {\bibinfo {volume} {111}},\ \bibinfo
  {pages} {073602} (\bibinfo {year} {2013})}\BibitemShut {NoStop}%
\bibitem [{\citenamefont {{Navarrete-Benlloch}}\ \emph
  {et~al.}(2014)\citenamefont {{Navarrete-Benlloch}}, \citenamefont
  {{Garc{\'{\i}}a-Ripoll}},\ and\ \citenamefont {{Porras}}}]{navarrete14prl}%
  \BibitemOpen
  \bibfield  {author} {\bibinfo {author} {\bibfnamefont {C.}~\bibnamefont
  {{Navarrete-Benlloch}}}, \bibinfo {author} {\bibfnamefont {J.~J.}\
  \bibnamefont {{Garc{\'{\i}}a-Ripoll}}}, \ and\ \bibinfo {author}
  {\bibfnamefont {D.}~\bibnamefont {{Porras}}},\ }\href {\doibase
  10.1103/PhysRevLett.113.193601} {\bibfield  {journal} {\bibinfo  {journal}
  {Phys. Rev. Lett.}\ }\textbf {\bibinfo {volume} {113}},\ \bibinfo {eid}
  {193601} (\bibinfo {year} {2014})}\BibitemShut {NoStop}%
\bibitem [{\citenamefont {Metelmann}\ and\ \citenamefont
  {T\"ureci}(2018)}]{Metelmann17}%
  \BibitemOpen
  \bibfield  {author} {\bibinfo {author} {\bibfnamefont {A.}~\bibnamefont
  {Metelmann}}\ and\ \bibinfo {author} {\bibfnamefont {H.~E.}\ \bibnamefont
  {T\"ureci}},\ }\href {\doibase 10.1103/PhysRevA.97.043833} {\bibfield
  {journal} {\bibinfo  {journal} {Phys. Rev. A}\ }\textbf {\bibinfo {volume}
  {97}},\ \bibinfo {pages} {043833} (\bibinfo {year} {2018})}\BibitemShut
  {NoStop}%
\bibitem [{\citenamefont {Quijandr\'{\i}a}\ \emph {et~al.}(2018)\citenamefont
  {Quijandr\'{\i}a}, \citenamefont {Naether}, \citenamefont {\"Ozdemir},
  \citenamefont {Nori},\ and\ \citenamefont {Zueco}}]{Quijandria18}%
  \BibitemOpen
  \bibfield  {author} {\bibinfo {author} {\bibfnamefont {F.}~\bibnamefont
  {Quijandr\'{\i}a}}, \bibinfo {author} {\bibfnamefont {U.}~\bibnamefont
  {Naether}}, \bibinfo {author} {\bibfnamefont {S.~K.}\ \bibnamefont
  {\"Ozdemir}}, \bibinfo {author} {\bibfnamefont {F.}~\bibnamefont {Nori}}, \
  and\ \bibinfo {author} {\bibfnamefont {D.}~\bibnamefont {Zueco}},\ }\href
  {\doibase 10.1103/PhysRevA.97.053846} {\bibfield  {journal} {\bibinfo
  {journal} {Phys. Rev. A}\ }\textbf {\bibinfo {volume} {97}},\ \bibinfo
  {pages} {053846} (\bibinfo {year} {2018})}\BibitemShut {NoStop}%
\bibitem [{\citenamefont {Cirac}\ \emph {et~al.}(1992)\citenamefont {Cirac},
  \citenamefont {Blatt}, \citenamefont {Zoller},\ and\ \citenamefont
  {Phillips}}]{cirac92}%
  \BibitemOpen
  \bibfield  {author} {\bibinfo {author} {\bibfnamefont {J.~I.}\ \bibnamefont
  {Cirac}}, \bibinfo {author} {\bibfnamefont {R.}~\bibnamefont {Blatt}},
  \bibinfo {author} {\bibfnamefont {P.}~\bibnamefont {Zoller}}, \ and\ \bibinfo
  {author} {\bibfnamefont {W.~D.}\ \bibnamefont {Phillips}},\ }\href {\doibase
  10.1103/PhysRevA.46.2668} {\bibfield  {journal} {\bibinfo  {journal} {Phys.
  Rev. A}\ }\textbf {\bibinfo {volume} {46}},\ \bibinfo {pages} {2668}
  (\bibinfo {year} {1992})}\BibitemShut {NoStop}%
\bibitem [{\citenamefont {Fitzpatrick}\ \emph {et~al.}(2017)\citenamefont
  {Fitzpatrick}, \citenamefont {Sundaresan}, \citenamefont {Li}, \citenamefont
  {Koch},\ and\ \citenamefont {Houck}}]{fitzpatrick17}%
  \BibitemOpen
  \bibfield  {author} {\bibinfo {author} {\bibfnamefont {M.}~\bibnamefont
  {Fitzpatrick}}, \bibinfo {author} {\bibfnamefont {N.~M.}\ \bibnamefont
  {Sundaresan}}, \bibinfo {author} {\bibfnamefont {A.~C.~Y.}\ \bibnamefont
  {Li}}, \bibinfo {author} {\bibfnamefont {J.}~\bibnamefont {Koch}}, \ and\
  \bibinfo {author} {\bibfnamefont {A.~A.}\ \bibnamefont {Houck}},\ }\href
  {\doibase 10.1103/PhysRevX.7.011016} {\bibfield  {journal} {\bibinfo
  {journal} {Phys. Rev. X}\ }\textbf {\bibinfo {volume} {7}},\ \bibinfo {pages}
  {011016} (\bibinfo {year} {2017})}\BibitemShut {NoStop}%
\bibitem [{\citenamefont {Fern\'andez-Lorenzo}\ \emph
  {et~al.}(2018)\citenamefont {Fern\'andez-Lorenzo}, \citenamefont
  {Dunningham},\ and\ \citenamefont {Porras}}]{Fernandez18}%
  \BibitemOpen
  \bibfield  {author} {\bibinfo {author} {\bibfnamefont {S.}~\bibnamefont
  {Fern\'andez-Lorenzo}}, \bibinfo {author} {\bibfnamefont {J.~A.}\
  \bibnamefont {Dunningham}}, \ and\ \bibinfo {author} {\bibfnamefont
  {D.}~\bibnamefont {Porras}},\ }\href {\doibase 10.1103/PhysRevA.97.023843}
  {\bibfield  {journal} {\bibinfo  {journal} {Phys. Rev. A}\ }\textbf {\bibinfo
  {volume} {97}},\ \bibinfo {pages} {023843} (\bibinfo {year}
  {2018})}\BibitemShut {NoStop}%
\bibitem [{\citenamefont {Nevado}\ \emph {et~al.}(2017)\citenamefont {Nevado},
  \citenamefont {Fern\'andez-Lorenzo},\ and\ \citenamefont
  {Porras}}]{Nevado17}%
  \BibitemOpen
  \bibfield  {author} {\bibinfo {author} {\bibfnamefont {P.}~\bibnamefont
  {Nevado}}, \bibinfo {author} {\bibfnamefont {S.}~\bibnamefont
  {Fern\'andez-Lorenzo}}, \ and\ \bibinfo {author} {\bibfnamefont
  {D.}~\bibnamefont {Porras}},\ }\href {\doibase
  10.1103/PhysRevLett.119.210401} {\bibfield  {journal} {\bibinfo  {journal}
  {Phys. Rev. Lett.}\ }\textbf {\bibinfo {volume} {119}},\ \bibinfo {pages}
  {210401} (\bibinfo {year} {2017})}\BibitemShut {NoStop}%
\end{thebibliography}%
	
	%
	
\end{document}